%% file: main.tex
\documentclass[twocolumn,tighten,astrosymb,trackchanges]{aastex631}
\input{affiliations}
\usepackage{CJK}
\usepackage{multirow}
\usepackage[caption=false]{subfig}
\newcommand\kms{~km~s$^{-1}$}
\newcommand\mch{$M_{\text{Ch}}$}
\newcommand\um{~$\mu$m}
\let\tablenum\relax
\usepackage{siunitx}
\usepackage{appendix}
\usepackage{amsmath,amsthm,amsfonts,amssymb,amscd}
\usepackage{booktabs}
\usepackage{tablefootnote}
\usepackage{apjfonts}
\usepackage{lipsum}

\definecolor{maroon}{rgb}{0.760,0.118,0.337}

\definecolor{darkaqua}{rgb}{0.0,0.45,0.65}

\def\msun{\hbox{\,$M_{\odot}$}}

\def\cm{\mbox{\,cm}}
\def\cm3{\mbox{\,cm$^{-3}$}}

\shorttitle{JWST Evidence for a White Dwarf Merger in SN~2022pul}
\shortauthors{Kwok et al.}
\submitjournal{\textit{The Astrophysical Journal}}
\begin{document}

\title{Ground-based and JWST Observations of SN~2022pul: \\ 
II. Evidence from Nebular Spectroscopy for a Violent Merger in a Peculiar Type Ia Supernova}

\correspondingauthor{Lindsey A. Kwok}
\email{lindsey.kwok@physics.rutgers.edu}

\input{authors}

\begin{abstract}

We present an analysis of ground-based and \textit{JWST} observations of SN~2022pul, a peculiar ``03fg-like" (or ``super-Chandrasekhar") Type Ia supernova (SN Ia), in the nebular phase at 338~d post explosion. Our combined spectrum continuously covers 0.4--14\um\ and includes the first mid-infrared spectrum of an 03fg-like SN~Ia. Compared to normal SN Ia 2021aefx, SN~2022pul exhibits a lower mean ionization state, asymmetric emission-line profiles, stronger emission from the intermediate-mass elements (IMEs) argon and calcium, weaker emission from iron-group elements (IGEs), and the first unambiguous detection of neon in a SN~Ia. Strong, broad, centrally peaked [\ion{Ne}{2}] line at 12.81\um\ was previously predicted as a hallmark of ``violent merger'' SN~Ia models, where dynamical interaction between two sub-\mch\ white dwarfs (WDs) causes disruption of the lower mass WD and detonation of the other. The violent merger scenario was already a leading hypothesis for 03fg-like SNe~Ia; in SN~2022pul it can explain the large-scale ejecta asymmetries seen between the IMEs and IGEs and the central location of narrow oxygen and broad neon. We modify extant models to add clumping of the ejecta to better reproduce the optical iron emission, and add mass in the innermost region ($< 2000$\kms) to account for the observed narrow [\ion{O}{1}]~$\lambda\lambda6300$, 6364 emission. A violent WD-WD merger explains many of the observations of SN~2022pul, and our results favor this model interpretation for the subclass of 03fg-like SN~Ia.

\end{abstract}


\section{Introduction \label{sec:intro}}
Supernovae (SNe) imprint important information about the composition, geometry, kinematics, and ionization of the ejected material in their late-time (nebular) emission-line spectra. In the nebular phase \citep[$\gtrsim 100$ days past maximum light;][]{Bowers1997, Branch2008, Silverman2013, Friesen2014, Black2016}, the ejecta have expanded and the opacity has dropped enough to see directly through even the innermost regions and allow forbidden lines to form. In white-dwarf (WD) supernovae, decay of radioactive $^{56}$Co (produced by the radioactive decay of $^{56}$Ni synthesized in the explosion) provides most of the energy that excites the low-density nebular ejecta, producing forbidden emission lines throughout the ejecta \citep{Axelrod1980,Nadyozhin1994, jerkstrand_spectra_2017}. These glowing ashes contain clues for determining the SN origin. Nebular-phase spectroscopy is therefore an important forensic tool for investigating the remains of exploded WDs, whose explosion channels are not well understood. 

The majority of WD SNe are classified as Type Ia supernovae (SNe~Ia); however, a growing number and variety of extreme or peculiar SN~Ia subtypes have been discovered \citep[for recent reviews, see][]{Taubenberger2017,Jha2019,Liu2023}. These peculiar SNe result from the thermonuclear explosion of a WD, but they generally do not follow the \citet{Phillips1993} relationship between luminosity and light-curve shape that makes normal SNe~Ia so useful for cosmological distance measurements \citep{Phillips1993, Benetti2005, Taubenberger2017, Jha2019}. No consensus has been reached concerning the possible and dominant pathways connecting progenitor systems, explosion mechanisms, and observed WD SN properties.

The origins of normal SNe~Ia are particularly elusive in part because variations between objects are relatively small and many of the models predict similar observational properties. By contrast, extreme and peculiar WD SNe exhibit distinguishing features that can make identification of their origins clearer \citep{Foley2013, Kromer2013, Hsiao2020}. Understanding the scenarios that produce peculiar WD SNe helps us understand the unusual and extreme ways WDs can explode while simultaneously narrowing the model space for normal SNe~Ia.

A rare type of peculiar WD SN characterized by high luminosities, broad light curves, \ion{C}{2} absorption at early times, relatively low ejecta velocities, and low ionization state nebular-phase spectra are ``03fg-like" (commonly called ``super-Chandrasekhar") SNe~Ia \citep{howell_type_2006, Hicken2007, Scalzo2010, Silverman2011, Taubenberger2013, Hsiao2020, Ashall2021, Srivastav2023}. Explaining the light curve of SN~2003fg, the class prototype, by decay of $^{56}$Ni alone requires a total mass above 1.4~\msun, the Chandrasekhar mass limit (\mch) for a nonrotating WD \citep{howell_type_2006}. A mass above this limit can be achieved either by additional support from differential rotation in a single WD \citep{Yoon05} or a merger of two WDs with total mass greater than \mch\ \citep{howell_type_2006, Hicken2007}. Alternative explanations that have been invoked to explain this diverse class of bright SNe~Ia include the ``core-degenerate" scenario, where a near-\mch\ C/O WD explodes within the C-rich envelope of an asymptotic giant branch (AGB) star \citep{Khokhlov1993, Hoeflich1996,Livio2003, Quimby2007, Kashi2011, Hsiao2020, Lu2021}, and the violent merger of two sub-\mch\ WDs \citep{Dimitriadis22, Srivastav2023, Siebert2023}. These potential avenues account for the excess luminosity and relatively low velocity of these objects by interaction with the AGB star envelope or C/O-rich circumstellar material (CSM) from the disrupted smaller WD, respectively. For more comprehensive reviews of 03fg-like SNe~Ia, see \citet{Taubenberger2017}, \citet{Ashall2021}, and Paper I \citep{Siebert2023b}.

In contrast to the classical WD merger scenario where the merger produces a single-object remnant that later explodes, the violent merger model proceeds via dynamical interaction between two sub-\mch\ WDs that causes disruption of the secondary (lower mass) WD and subsequent detonation of the primary (higher mass) WD \citep[for a review, see][]{Pakmor2017}. The primary WD is burned first and its ejecta expand beyond the secondary by the time that the secondary is burned $\sim 1$~s later and the ejecta of the secondary expand into the center of the ejecta of the primary. Thus, the intermediate-mass elements (IMEs) produced by burning of the secondary will be centrally located within the ejecta. Depending on the mass ratio, mass of the primary WD, density of the secondary WD at the time of detonation, and composition of both WDs, a fairly wide range of SN properties can be produced \citep{Pakmor2011, Pakmor2012, Kromer2013, Pakmor2017}. In general, violent mergers are asymmetric scenarios which can result in enhanced production of IMEs from burning of the secondary, lower levels of iron-group elements (IGEs) from burning of the sub-\mch\ primary, centrally located C/O/Ne/Mg and IMEs, and a low ionization state at late times due to high central ejecta density. 

Nebular spectra can reveal all of these properties. This information can be difficult to disentangle, however, owing to significant line crowding and blending at optical wavelengths. Spectroscopy with \textit{JWST} in the near-infrared (NIR, which extends beyond the NIR wavelength range accessible from the ground) and mid-infrared (MIR) grants access to emission lines from several species of IMEs, and the lines are comparatively isolated so individual line profiles can be distinguished. These NIR and MIR spectral features help clarify the information contained in the optical and ground-based NIR spectra, aiding in discriminating between progenitor systems and explosion mechanisms. Observations of SN~2021aefx, the first SN~Ia with \textit{JWST} NIR$+$MIR nebular spectra, revealed strong, stable Ni indicative of a high-mass WD progenitor and evidence for stratified ejecta resulting from some type of detonation \citep{Kwok2023}. Further modeling found SN~2021aefx to be consistent with an off-center delayed detonation of a near-\mch\ mass WD \citep{DerKacy2023}.

\begin{figure*}
    \centering
    \includegraphics[width=1.0\textwidth]{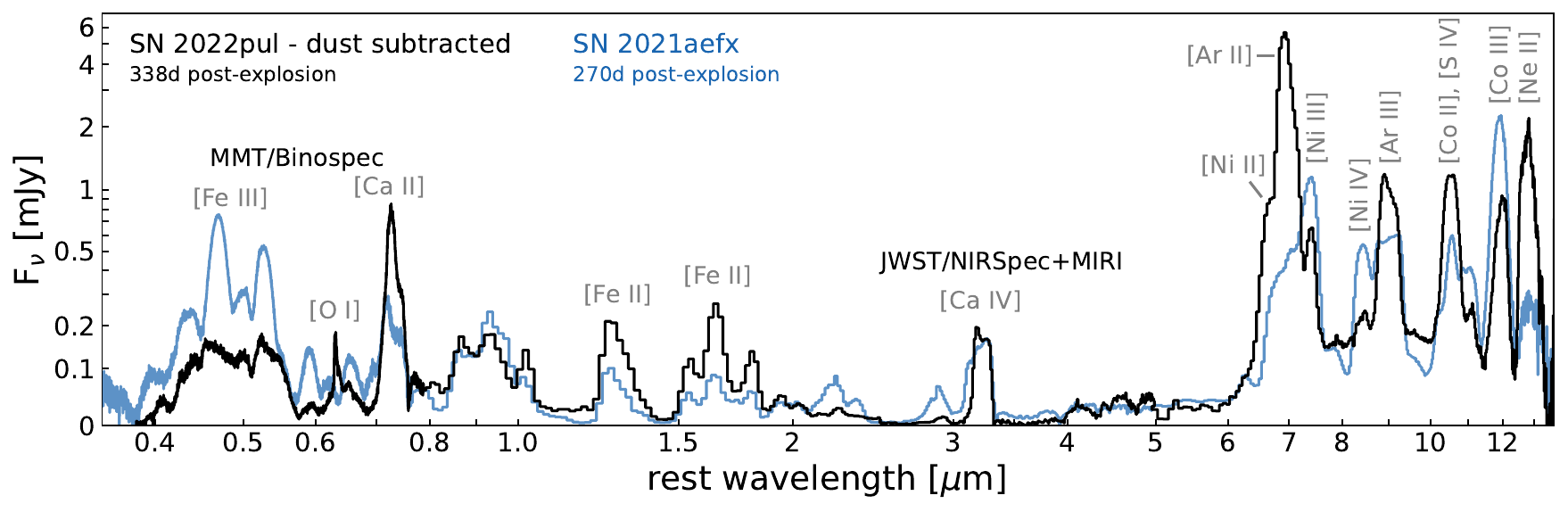}
    \caption{Full optical$+$NIR$+$MIR comparison between the dust-continuum-subtracted spectrum of SN 2022pul at 338 rest-frame days post-explosion ($t_{\text{exp}}=$ MJD 59785.3) at $d=16$~Mpc, and SN 2021aefx at 270~d rest-frame days post-explosion at $d=18$ Mpc from \citet{Kwok2023}. The optical spectrum of SN~2022pul is from MMT/Binospec at a similar epoch of 332~d post-explosion and the NIR$+$MIR spectrum is from \textit{JWST}/NIRSpec$+$MIRI. The prominent stable Ni emission lines in SN 2021aefx are weaker (relatively) in SN 2022pul, as is the [\ion{Co}{3}] 11.88\um\ line. In SN~2022pul, the singly ionized emission lines are relatively strong compared with the higher ionization states. A strong [\ion{Ne}{2}] feature at 12.81\um\ is observed in SN 2022pul. The flux axis uses a non-linear (arcsinh) scale to better show all the features across a wide range of wavelength and $F_\nu$.}
    \label{fig:full_spec}
\end{figure*}

We obtained a \textit{JWST} NIR$+$MIR nebular spectrum of SN 2022pul, a peculiar 03fg-like SN~Ia (\textit{JWST} Cycle 1 GO 2072; PI S.~W.~Jha). Our \textit{JWST} spectrum of SN~2022pul is the first MIR spectrum of an 03fg-like SN~Ia, and it exhibits a strong thermal ($T \approx 500$~K) dust continuum \citep[analyzed in Paper III;][]{Johansson2023}. It also displays several differences from the optical--MIR spectrum of SN 2021aefx. Additional details about SN~2022pul and its unique photometric and spectroscopic evolution can be found in Paper I \citep{Siebert2023b}.

Here, we present analysis of the dust-continuum-subtracted optical through MIR nebular-phase spectrum of SN 2022pul at 338~d post explosion (this assumes that the dust is external, see \citealt{Johansson2023} for a discussion). In \autoref{sec:spec_analysis} we identify NIR and MIR emission lines and highlight distinctive spectral properties in comparison to SN~2021aefx. \autoref{sec:line_profiles} presents emission-line-profile fits to the dominant spectral features and their implications on the distribution of material in the ejecta. In \autoref{sec:model} we compare SN~2022pul to nebular spectroscopic predictions \citep{Blondin2023} of the violent merger model from \citet{Pakmor2012}. We discuss the implications of our results and conclude in \autoref{sec:conclusions} that SN~2022pul was most likely produced by the violent merger of two WDs.

\section{Spectral Analysis \label{sec:spec_analysis}}

The continuous optical$+$NIR$+$MIR spectrum of SN~2022pul in the nebular phase at 338 rest-frame days post-explosion (estimated explosion date MJD 59785.3) exhibits many unique properties, including a clear NIR$+$MIR dust continuum that closely resembles a blackbody \citep[see also][]{Siebert2023b}. Here, we focus on analysis of the spectral lines, while a detailed analysis of the dust properties is given by \citet{Johansson2023}. We remove the dust contribution to the spectrum by subtracting off a blackbody of $T = 500$~K from the 2.5--14\um\ region, and use this dust-subtracted spectrum throughout this work. We do not observe other spectral signs of dust such as CO or SiO emission (although line overlap may complicate clear identification of dust extinction in the optical). For details of the observations and data reduction of the spectra used in this work, see \citet{Siebert2023b}. 

Shown in \autoref{fig:full_spec}, several highlights of SN~2022pul's combined nebular spectrum include a low mean ionization state, strong emission from IMEs, relatively weak emission from IGEs, highly asymmetric line profiles, narrow [\ion{O}{1}]~$\lambda\lambda6300,~6364$ and remarkably strong [\ion{Ca}{2}]~$\lambda\lambda7291,~7324$ in the optical spectrum, detection of [\ion{S}{4}] at 10.51\um, and the first unambiguous detection of [\ion{Ne}{2}] at 12.81\um\ in an SN~Ia.

\begin{figure*}
    \centering
    \includegraphics[width=1.0\textwidth]{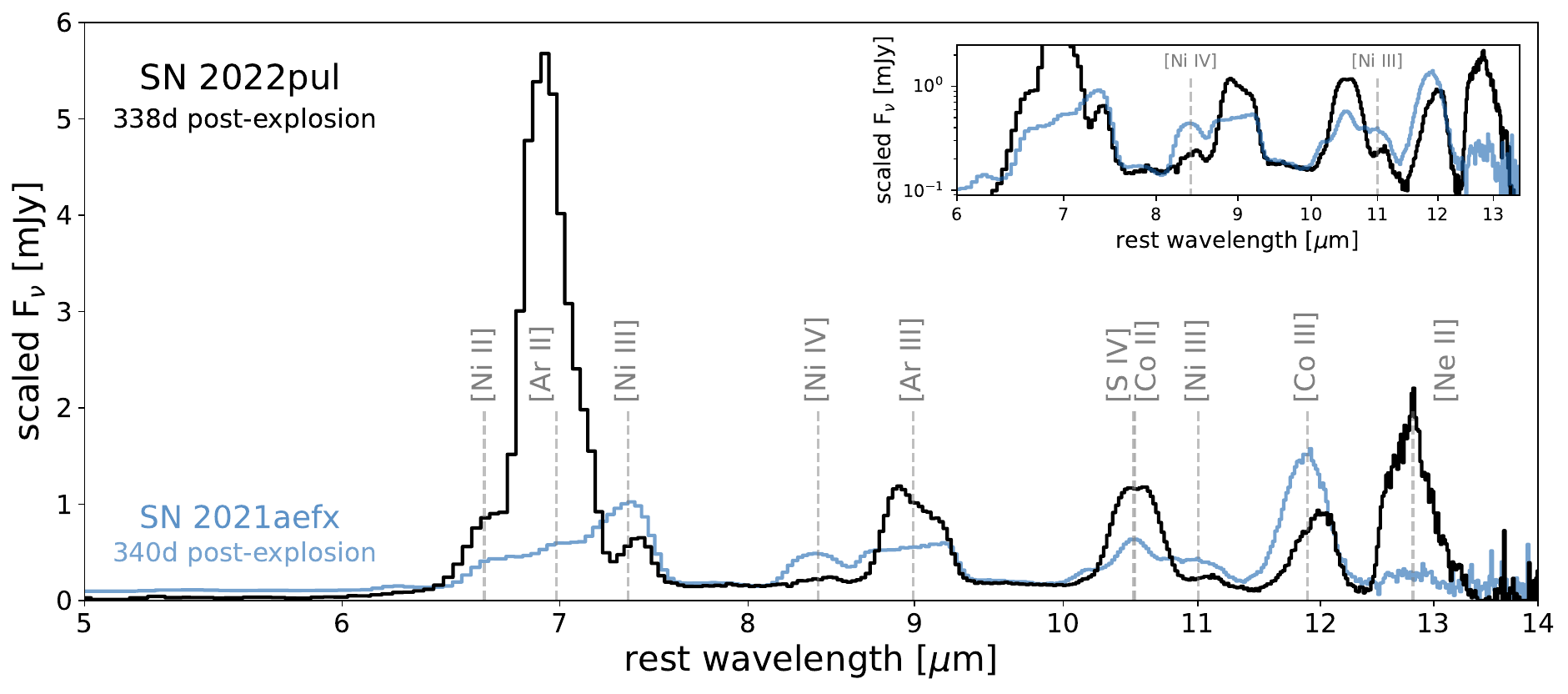}
    \caption{Comparison between the dust-continuum-subtracted MIR spectrum of SN 2022pul at 338 rest-frame days post-explosion (MJD 59785.3) and the MIR spectrum of SN 2021aefx at 340d rest-frame days post-explosion from \citet{DerKacy2023} scaled to the distance of SN~2022pul (16 Mpc). The prominent stable $^{58}$Ni and Co emission lines in SN~2021aefx are somewhat weaker and have shifted to lower mean ionization state in SN~2022pul. The [\ion{Ar}{2}] 6.98\um\ line dominates the SN~2022pul spectrum, with strong [\ion{Ar}{3}] 8.99\um\ and [\ion{S}{4}] 10.51\um\ as well. [\ion{Ne}{2}] at 12.81\um\ is strong, broad, and centrally peaked in SN~2022pul. All line profiles are considerably asymmetric. An inset is provided on a log scale to show the very weak [Ni IV] 8.41\um\ and [\ion{Ni}{3}] 11.00\um\ lines.}
    \label{fig:MIR_spec}
\end{figure*}

\subsection{Spectroscopic Comparison with SN 2021aefx}

SN~2021aefx, a spectroscopically and photometrically regular SN~Ia near maximum light, is currently the only other SN Ia with a published continuous spectrum from 0.3\um\ to 14\um\ in the nebular phase \citep[270~d post-explosion;][]{Kwok2023}, making it a natural comparison object for the peculiar ``super-Chandra'' 03fg-like SN~2022pul. After subtracting the dust continuum, SN~2022pul has a greater fraction of total flux in the NIR and MIR, and a smaller fraction of optical flux than SN~2021aefx. The light curve of SN~2022pul shows that at late times, the \textit{I}-band flux exceeds the flux in other bands (probably due to the strong [\ion{Ca}{2}]~$\lambda\lambda7291,~7324$ doublet) and declines more slowly than typical SNe~Ia \citep{Siebert2023b}. This is consistent with findings from \citet{Ashall2021} that 03fg-like SNe Ia are brighter and decline more slowly in the NIR than normal SNe Ia.

This redder color may be due in part to the comparatively low overall ionization state of SN~2022pul, where additional flux from lines of lower ionization falls in the infrared. The flux from optical [\ion{Fe}{3}] around 5000~\AA\ is greatly diminished relative to most SNe~Ia \citep{Siebert2023b}; similarly, the \textit{JWST} NIRSpec spectrum reveals that the NIR [\ion{Fe}{3}] lines around 2.2 and 2.9\um\ are weak while the NIR [\ion{Fe}{2}] lines around 1.3 and 1.6\um\ are strong compared with SN~2021aefx. In the MIR, we see the same ionization trend in the Ni, Ar, and Co where the ionization is lower in SN~2022pul than in SN~2021aefx. In agreement with the ionization inferred from the MIR, the strength of the [\ion{Ca}{2}]~$\lambda\lambda7291,~7324$ doublet in SN 2022pul is remarkable \citep{Siebert2023b}, whereas in SN~2021aefx the spectral complex near 7300~\AA\ was attributed to [\ion{Fe}{2}] and [\ion{Ni}{2}] with little to no contribution from [\ion{Ca}{2}] \citep{Kwok2023} --- yet they have similar strengths of [\ion{Ca}{4}] near 3.2\um. This shift to a lower mean ionization for SN~2022pul suggests a larger density or mass of the ejecta, but the presence of \ion{Ca}{4} also indicates that some higher ionization bubbles are present in both, discussed further in \autoref{sec:model}.

The NIR does not display lines other than those present in SN 2021aefx, though the [\ion{Fe}{3}] is weaker and the three [\ion{Fe}{2}] features in the range 4--5\um\ are stronger. \citet{Blondin2023} showed that the broad feature near 3.2\um\ is dominated by [\ion{Ca}{4}] 3.21\um, with weaker contributions from [\ion{Fe}{3}] 3.23\um\ and [\ion{Ni}{1}] 3.12\um. The NIRSpec observation of SN~2022pul further confirms this broad line around 3.2\um\ as [\ion{Ca}{4}] 3.21\um\ because the [\ion{Fe}{3}] and Ni emission are weaker in SN~2022pul than in SN~2021aefx, but the 3.2\um\ feature has a similar strength. Several unique features in the MIRI/LRS spectrum of SN~2022pul are shown in \autoref{fig:MIR_spec} and described below.

\subsection{[Ne II] emission}  % don't use \ion in section heading
In addition to all of the NIR and MIR lines present in SN~2021aefx (line identifications are given by \citealt{Kwok2023}), SN~2022pul exhibits a strong, broad, centrally peaked feature around 12.8\um\ that we identify as [\ion{Ne}{2}] 12.81\um\ (see \autoref{fig:MIR_spec}). Predicted by \citet{Blondin2023}, the presence of strong, centrally peaked [\ion{Ne}{2}] 12.81\um\ is a distinguishing feature of a violent WD--WD merger not present in any of the other models considered in their study. Ne is produced centrally in a violent WD--WD merger because it is a product of burning the less dense secondary WD, whose ejecta expands into the center of the primary WD's ejecta after the primary has burned and expanded. The observed [\ion{Ne}{2}] 12.81\um\ line is broad, asymmetric, and dissimilar to other line-profile shapes seen in SN~2022pul and other SNe~Ia with MIR spectra.

\subsection{Strong Intermediate-Mass Elements}
One of the major differences between the MIR spectra of SN~2022pul and SN~2021aefx is the strength of the IME (hereafter defined as S, Ar, and Ca) emission lines. Whereas in SN~2021aefx, [\ion{Ar}{2}] 6.98\um\ was blended out by the neighboring Ni emission lines such that its shape was uncertain, [\ion{Ar}{2}] 6.98\um\ is the dominant emission line in the MIR spectrum of SN~2022pul. Its peak flux is blueshifted by $\sim 2000$\kms, and is over twice as strong in $F_\nu$ as [\ion{Ne}{2}] 12.81\um, the next strongest MIR emission line. Similarly, [\ion{Ca}{2}]~$\lambda\lambda7291,~7324$ dominates the optical spectrum.

The [\ion{Ar}{3}] 8.99\um\ emission line is also quite strong in SN~2022pul: roughly three times stronger in $F_\nu$ than in SN~2021aefx at a very similar time post-explosion (see \autoref{fig:MIR_spec}). The [\ion{Ar}{3}] 8.99\um\ line continues to exhibit a boxy, flat-topped shape, with a similar width, but the slanted top is much steeper. This steep slant can be produced by an asymmetric shell of [\ion{Ar}{3}] emission, discussed further in \autoref{sec:line_profiles}.

The feature at 10.5\um\ in SN~2022pul results from a combination of [\ion{S}{4}] 10.51\um\ and [\ion{Co}{2}] 10.52\um\ emission. The strength and flat shape of the feature at 10.5\um\ cannot be explained by [\ion{Co}{2}] 10.52\um\ alone, as is further discussed in \autoref{sec:line_profiles}, constituting a firm detection of [\ion{S}{4}]. \citet{Kwok2023} did not invoke a contribution from [S~IV] 10.51\um\ in SN~2021aefx; however, placed in context with SN~2022pul, the ``shoulder" feature starting at 10\um\ in SN~2021aefx might actually be [\ion{Si}{4}] with a similar boxy shape as the [\ion{Ar}{3}] but with about half the strength. Re-evaluation of [\ion{S}{4}] in SN~2021aefx will be included in future work.

All of the IME emission lines in SN~2022pul are strong relative to SN~2021aefx, and their peak emission is asymmetric and blueshifted.

\subsection{Weak Iron-group Elements}

In contrast to the strength of the IMEs, the IGEs (hereafter Fe, Co, and Ni) appear somewhat weaker in the MIR than in SN~2021aefx. Part or all of this apparent decrease in IGEs may be due to the shift to lower ionization. For example, the [\ion{Ni}{2}] 6.64\um\ line is much stronger in SN~2022pul, but the [\ion{Ni}{3}] lines at 7.35 and 11.00\um\ are several times weaker, and the [\ion{Ni}{4}] 8.41\um\ line is so weak that it is difficult to see without an arcsinh or log scaling of the flux. This points toward some of the flux in the higher ionization IGE emission lines being shifted to the singly ionized emission lines.

The isolated [\ion{Co}{3}] 11.89\um\ line shows a distinctly asymmetric shape whose peak is redshifted by $\sim 3500$\kms. Compared with SN~2021aefx, the [\ion{Co}{3}] 11.89\um\ line is somewhat narrower. As will be shown and discussed in \autoref{sec:line_profiles}, the asymmetric redshifted shape of the [\ion{Co}{3}] 11.89\um\ line profile is consistent with all of the other IGE ions throughout the NIR and MIR, as well as the redder side of the optical.

\section{Emission-Line Profiles \label{sec:line_profiles}}

\begin{figure}
    \centering
    \includegraphics[width=0.5\textwidth]{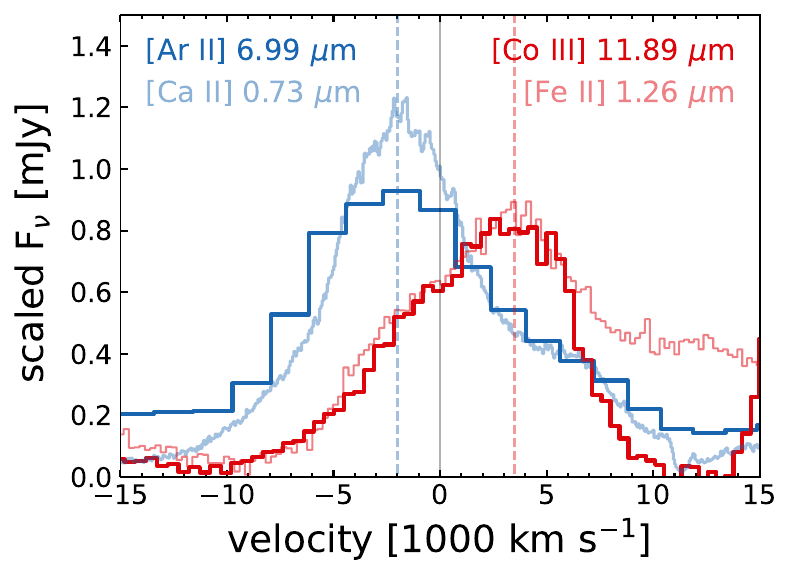}
    \caption{[\ion{Ar}{2}] and [\ion{Ca}{2}] emission-line profiles in velocity space compared with [\ion{Co}{3}] and [\ion{Fe}{2}]. The shapes of the [\ion{Ar}{2}] and [\ion{Ca}{2}] are quite consistent with each other, and the shapes of [\ion{Co}{3}] and [\ion{Fe}{2}] are nearly identical (taking into account the neighboring [\ion{Fe}{2}] lines that create the red tail on the [\ion{Fe}{2}]~1.26\um\ feature). Notably, the Ar and Ca (IMEs) have a blueshifted peak with a more extended red tail, while the Co and Fe (IGEs) display an opposite trend, with a redshifted peak and a more extended blue tail.}
    \label{fig:IGE_IME_comp}
\end{figure}

Nebular emission-line profiles are important indicators of the geometry of the SN ejecta because they imprint the flux at all projected line-of-sight velocities (and therefore projected radius) through the ejecta. The shape of the lines arises from a combination of the density and excitation of the elements within the ejecta \citep[for a review, see][]{jerkstrand_spectra_2017}. In SN~2022pul, we find evidence in the highly asymmetric line profiles for distinct distributions of the IGEs (Fe, Co, Ni), IMEs (S, Ar, Ca), O, and Ne within the ejecta. Unfortunately, it is not possible to determine the full three-dimensional (3D) geometry that gives rise to each profile owing to projection effects and the possibility of bubbles, mushroom shapes, clumps, or other such non-spherically symmetric distributions. However, by fitting line profiles throughout the optical, NIR, and MIR, we can determine the composition within slabs perpendicular to the line of sight, which resolves some of the geometrical layout. In this section we fit line profiles consistently across the optical+NIR+MIR and discuss their implications. 

For our line-profile fitting, we need relatively high resolution to discern line-profile shapes, so we combine an optical MMT/Binospec spectrum at 332~d post-explosion, a NIR Keck/NIRES spectrum at 316~d post-explosion (the nearest-phase ground-based NIR spectrum available), and the \textit{JWST}/NIRSpec$+$MIRI spectrum beyond 1.83\um\ at 338~d post-explosion \citep{Siebert2023b}. Despite the difference in phase, we use the Keck/NIRES spectrum (rebinned to increase the signal-to-noise ratio per pixel) instead of the \textit{JWST}/NIRSpec spectrum in the range 1.0--1.83\um\ owing to the much higher resolution. In addition to removing the dust contribution, we also remove all remaining underlying MIR continuum in the \textit{JWST} data \citep[see][for discussion of the MIR continuum]{Johansson2023}.

The optical, NIR, and MIR spectra of SN~2022pul are all essential to this work: the higher resolution of the ground-based optical and NIR spectra provides clarity for several key line-profile shapes, and the lower resolution \textit{JWST} data contain isolated lines and emission from IMEs. These properties of the data are complementary for interpreting the line profiles of SN~2022pul.

\subsection{Line-Fitting Procedure}
Following the general approach of nebular line fitting from \citet{Maguire2018}, \citet{Flors2018, Flors2020}, and \citet{Kwok2023}, we model the observed spectrum as a superposition of line profiles for all reasonably strong contributing lines. Restrictions on the fitting are enforced such that all lines of the same ion have the same profile shape parameters and kinematic offset from the central wavelength. We note similarities between the nebular spectrum of SN~2022pul and the nebular spectrum of the violent WD--WD merger model of \citet{Pakmor2012} computed by \citet{Blondin2023} (their MERGER model). Thus, we compute this model out to the same phase as SN~2022pul (338~d post-explosion) and we fix the relative line strengths of each contributing line in our fits to the relative line strengths in the model. This allows us to base the relative line strengths on temperatures and densities similar to those in SN~2022pul. Details of and comparisons to the merger model are presented in \autoref{sec:model}. 

While the relative line strengths within a given ion are fixed by the MERGER model at 338~d, we allow the relative amplitudes between each ion to vary as fitting parameters. We fit the optical through MIR and include the following ions: \ion{Fe}{1}, \ion{Fe}{2}, \ion{Fe}{3}, \ion{Co}{2}, \ion{Co}{3}, \ion{Ni}{2}, \ion{Ni}{3}, Ni IV, \ion{S}{4}, \ion{Ca}{2}, \ion{Ca}{4}, \ion{Ar}{2}, \ion{Ar}{3}, \ion{O}{1}, and \ion{Ne}{2}. Additionally, to account for asymmetry in SN~2022pul, we fit each ion with non-Gaussian line profiles, further described below.

The resolution of the NIRSpec/Prism and MIRI/LRS data introduces significant uncertainties into the fits for parameters that measure width. For features beyond 1.8\um, these resolution uncertainties are on the order of $1200$\kms\ near 3\um, $2000$\kms\ near 6\um, and $500$\kms\ near 12\um. Uncertainties from the fitting were calculated using a bootstrap resampling method. Uncertainties in width parameters such as the full width at half-maximum intensity (FWHM) and inner-shell velocity were obtained by adding the resolution and fitting uncertainties in quadrature.

\begin{figure*}
    \centering
    \includegraphics[width=\textwidth]{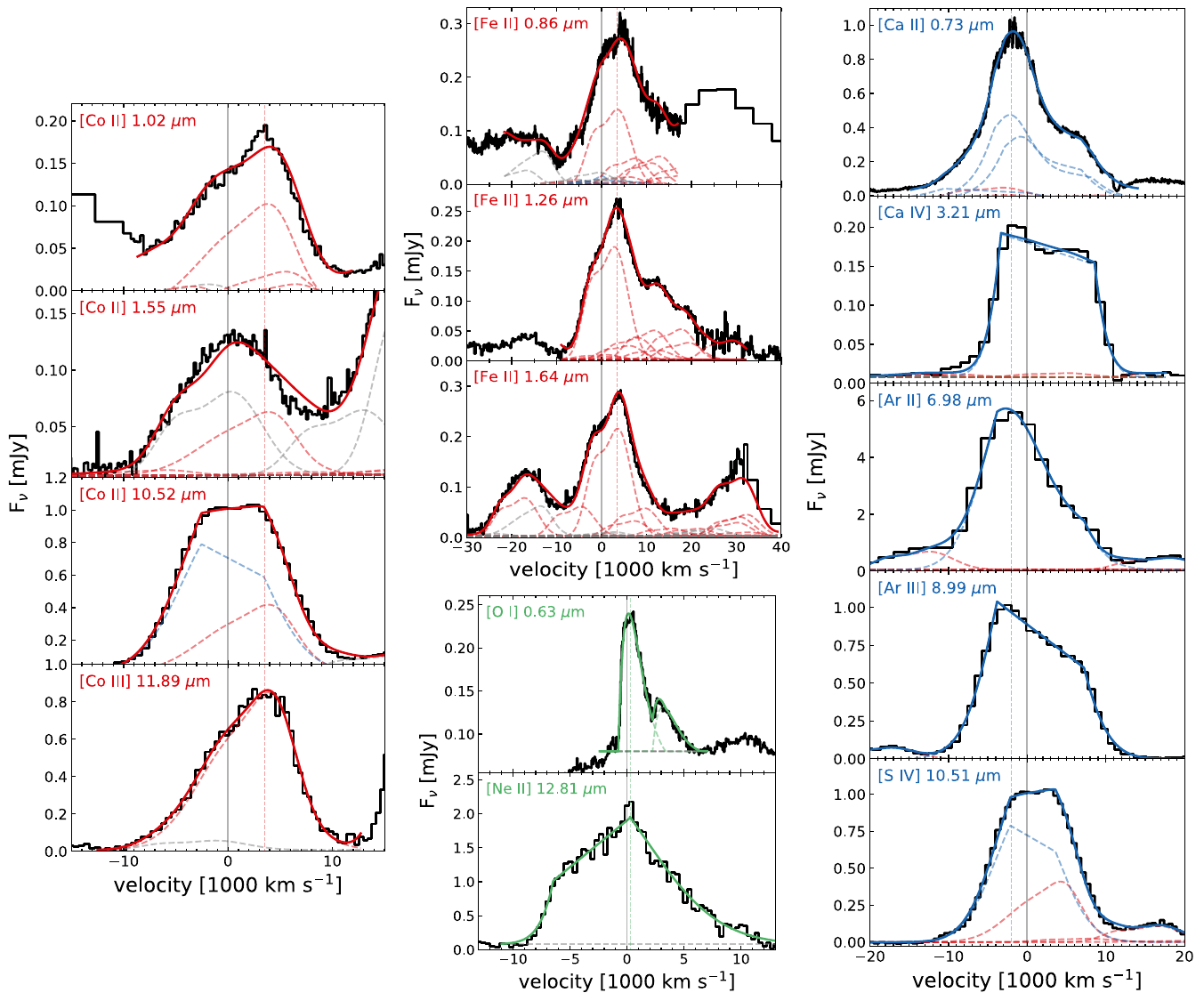}
    \caption{\textit{Left:} Fits to Co emission lines. The vertical dashed red line shows that the peak of the IGE distributions is highly redshifted at $\sim3500$\kms. Neighboring IME lines are dashed blue lines while neighboring Fe lines are dashed gray lines. \textit{Center top:} Similar to the Co fits, but for Fe (neighboring Co lines are dashed grey). \textit{Right:} Fits to the IME emission lines. Neighboring IGE lines are shown in dashed red. The vertical dashed blue line shows the peak of the IME distributions, which is highly blueshifted at $\sim 2000$\kms. \textit{Center bottom:} Fits to the optical [\ion{O}{1}]~$\lambda\lambda6300,~6364$ doublet and the MIR [\ion{Ne}{2}] 12.81\um\ line. The [\ion{O}{1}] is very central and narrow. The [\ion{Ne}{2}] peaks at the same location as the [\ion{O}{1}] (vertical dashed green line); however, it has an additional Gaussian-like red wing and a flat-topped-like blue wing slanted in the direction opposite that of the IMEs.}
    \label{fig:all_fits}
\end{figure*}

\begin{figure}
    \centering
    \includegraphics[width=0.4\textwidth]{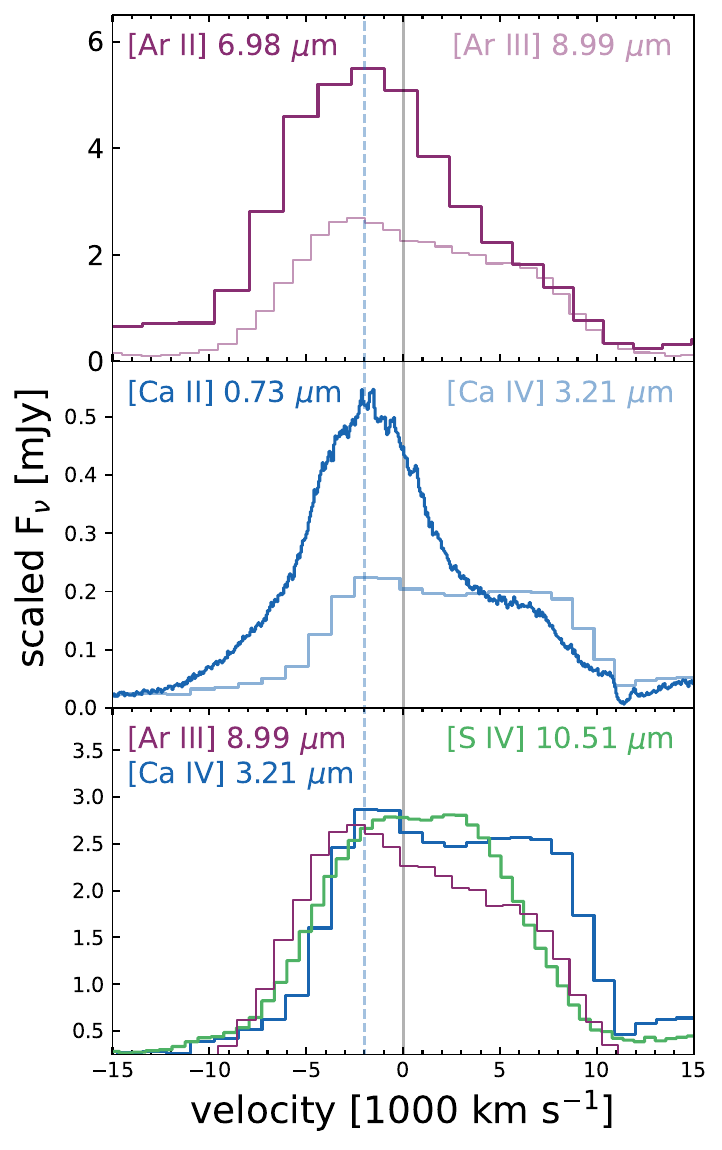}
    \caption{Comparison of the [\ion{Ar}{2}] 6.98\um\ and [\ion{Ar}{3}] 8.99\um\ lines, as well as the optical [\ion{Ca}{2}]~$\lambda\lambda7291,~7324$ and NIR [\ion{Ca}{4}] 3.21\um\ lines. The lower ionization states show a more prominent blueshifted peak (vertical dashed blue line), while the higher ionization states exhibit more of an asymmetric, slanted flat-topped shape. The bottom panel shows a comparison between the high ionization states of the IMEs. The [\ion{S}{4}] 10.51\um\ appears more flat-topped owing to the neighboring [\ion{Co}{2}] 10.52\um\ line peaked toward the red. The different profile shapes between ionization states imply ionization stratification.}
    \label{fig:IMEs_ionization}
\end{figure}

\subsection{Iron-Group Elements: Fe, Co, Ni}

Starting with the isolated and well-resolved [\ion{Co}{3}] 11.89\um\ line, we approximate the shape of the line profile by the sum of two Gaussians with FWHM $= 5400 \pm 1900$\kms\ and kinematic offset $v_{\text{offset}}= 4600 \pm 2300$\kms, and FWHM $= 8000 \pm 2200$\kms\ and kinematic offset $v_{\text{offset}} = 1400 \pm 1800$\kms. As shown in \autoref{fig:IGE_IME_comp}, the [\ion{Fe}{2}] 1.26\um\ line has a shape very similar (taking into account the neighboring [\ion{Fe}{2}] lines that create its red tail) to that of [\ion{Co}{3}] 11.89\um, so we fit all the [\ion{Fe}{2}] lines across 1.20--1.83\um\ with a sum of two Gaussians and obtain a similar, but slightly more peaked, profile for [\ion{Fe}{2}] with FWHM $=4700 \pm 1200$\kms\ and kinematic offset $v_{\text{offset}}=3800 \pm 500$\kms, and FWHM $= 7400 \pm 2100$\kms\ and kinematic offset $v_{\text{offset}}= -600 \pm 1300$\kms\ (see \autoref{fig:all_fits} for a visualization of a double Gaussian profile). 

A possible explanation for why Fe and Co do not exhibit exactly the same profiles is that while Co mostly traces the original $^{56}$Ni profile, Fe is more widely distributed because it is partially primordial. In the regions dominated by the IMEs and O$+$Ne, Fe may be a significant coolant since primordial Fe is $\sim400$ times more abundant than Co. Additionally, or alternatively, the Co may be less sharply peaked as this may be the region corresponding to partial burning in quasi nuclear statistical equilibrium (qNSE) \citep{DerKacy2023}.

Although the MIR isolated Ni lines are weak and low-resolution, [\ion{Ni}{2}], [\ion{Ni}{3}], and [\ion{Ni}{4}] all exhibit a very similar line profile shape. Likewise, the [\ion{Co}{2}] lines have a similar shape as [\ion{Co}{3}] and the [\ion{Fe}{3}] lines have a similar shape as [\ion{Fe}{2}]. Thus, to reduce degeneracy and make fitting many ions (and therefore parameters) more tractable, we use the same profile shape parameters (excluding amplitude) for all ionization states of the same element for Co, Fe, and Ni. We find that the Ni lines are slightly better fit by the Fe parameters than by Co, so we use the same profile-shape parameters as [\ion{Fe}{2}] for all Ni lines. In principle, we would not necessarily expect different ionization states of the same element to arise from the same emitting regions; however, \autoref{fig:all_fits} shows that our constraints on these profile fits can reproduce the emission features very well from the optical through the MIR. This suggests that there is not significant ionization stratification in the IGEs (i.e., emission from different ionization states is coming from similar ejecta regions). Furthermore, the IGE lines all peak with a redshift of $\sim 3500$\kms; the consistency of these profiles indicates that the IGEs are present in roughly the same ejecta regions.

\subsection{Intermediate-Mass Elements: S, Ar, Ca}

\begin{figure*}
    \centering
    \includegraphics[width=0.7\textwidth]{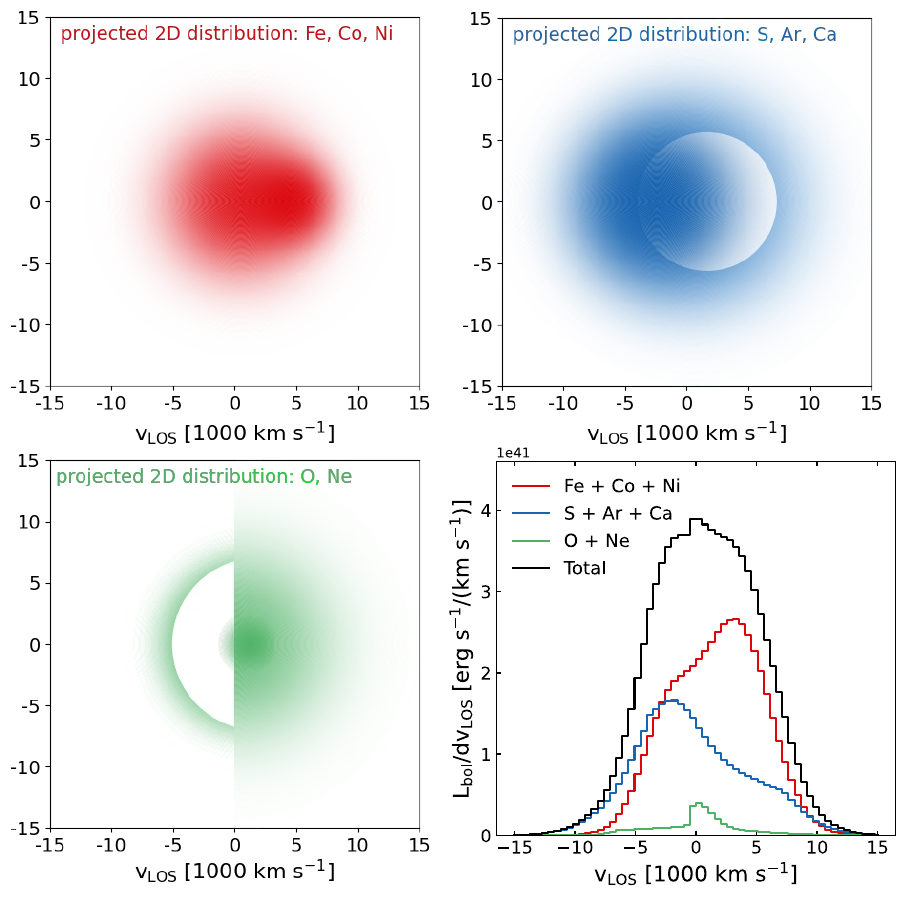}
    \caption{\textit{Top left, top right, bottom left}: Cartoon projected 2D geometric interpretations of the IGE, IME, and O$+$Ne line profiles, assuming axial symmetry. \textit{Bottom right:} Bolometric luminosity per projected velocity bin (500\kms) along the line of sight for the IGEs (Fe, Co, Ni), the IMEs (S, Ar, Ca), and O$+$Ne. There are clear differences in the overall line profiles from each element group, indicating distinct geometric distributions of material. This bolometric luminosity plot, derived from the spectral line profiles of the data (independent of the cartoon models), gives the true relative contributions of the different element groups without suggesting microscopic mixing.}
    \label{fig:cartoon_distributions}
\end{figure*}

The emission from IMEs in SN~2022pul is very strong and displays more line-profile variation than the IGEs. Despite the variation, the IME emission-line profiles share similar shapes and all exhibit a blueshifted peak at $\sim -2000$\kms, the opposite direction as the IGEs. They also exhibit ionization stratification, shown in \autoref{fig:IMEs_ionization}. The higher ionization states of [\ion{S}{4}], [\ion{Ar}{3}], and [\ion{Ca}{4}] all have a generally asymmetric boxy shape, whereas the lower ionization [\ion{Ar}{2}] and [\ion{Ca}{2}] appear to have an additional blueshifted, Gaussian-like component. We choose to fit the Ca II and Ar II features with the sum of a Gaussian and a slanted flat-top profile. Physically this could mean that the central, denser regions of the ejecta have lower overall ionization, requiring this additional Gaussian component (see \autoref{fig:cartoon_distributions} for a visualization of an asymmetric shell and additional Gaussian component profile).

The right panel of \autoref{fig:all_fits} shows our best-fit line profiles for the IMEs. The [\ion{Ar}{3}] 8.99\um\ line is isolated and exhibits a clear slanted flat-topped shape with Gaussian wings. This flat-topped Gaussian profile can arise from a thick shell of [\ion{Ar}{3}] emission, where one side of the shell is thicker than the other, causing the slant. The viewing angle of this asymmetric shell also contributes to the observed slant \citep{DerKacy2023}. The parameters for our [\ion{Ar}{3}] 8.99\um\ fit are FWHM $= 10,300 \pm 1100$\kms\ for the Gaussian wings, overall kinematic offset $v_{\text{offset}}=700 \pm 300$\kms, inner-shell velocity $v_{\text{min}}=5400 \pm 900$\kms, and offset of the shell thickness $v_{\text{offset, shell}}=\,-1000 \pm 200$\kms.

As described above, we fix the fit parameters (excluding amplitude) for [\ion{Co}{2}]~10.52\um\ and [\ion{Ni}{3}]~11.00\um, and then we fit the remaining contribution to the feature at 10.5\um\ as the [\ion{S}{4}]~10.51\um\ line. This yields a slanted boxy profile with Gaussian wings for [\ion{S}{4}], with a similar width and slope as [\ion{Ar}{3}], but a narrower inner-shell radius. Its parameters are FWHM $=\,8200 \pm 800 $\kms, kinematic offset $v_{\text{offset}}=\,-1000 \pm 200$\kms, inner-shell velocity $v_{\text{min}}=\,2500 \pm 700$\kms, and offset of the shell thickness $v_{\text{offset, shell}}=\,-1300 \pm 300$\kms. 

When we assume the peak of the [\ion{Co}{2}]~10.52\um\ line is redshifted (like all the relatively isolated IGE ions consistently display), we recover a blueshifted peak for [\ion{S}{4}] that agrees with the other IMEs. Because the IGEs are redshifted and the IMEs are blueshifted, we are able to distinguish the contributions to the 10.5\um\ feature from both [\ion{Co}{2}]~10.52\um\ and [\ion{S}{4}]~10.51\um. We do not attribute the full flat-topped, boxy shape of the feature at 10.5\um\ to [\ion{S}{4}]~10.51\um\ alone because the neighboring [\ion{Co}{3}]~11.89\um\ line is quite strong, and SN~2022pul has a low mean ionization state, so we do not expect the contribution from [\ion{Co}{2}]~10.52\um\ to be negligible.

The [\ion{Ca}{4}]~3.21\um\ line can also be fit reasonably well by an asymmetric boxy profile with slightly larger width than [\ion{Ar}{3}] and shallower slope. Its parameters are FWHM $=\,7500 \pm 1300 $\kms, kinematic offset $v_{\text{offset}}=\,-2400 \pm 200$\kms, inner-shell velocity $v_{\text{min}}=\,5900 \pm 1100$\kms, and offset of the shell thickness $v_{\text{offset, shell}}=\,-200 \pm 200$\kms. Simultaneously fitting the nearby [\ion{Fe}{3}] around 2.9\um, we constrain the contribution to the [\ion{Ca}{4}] profile from [\ion{Fe}{3}]~3.23\um\ to be very minimal. Thus, most of the flux in this feature is from [\ion{Ca}{4}], though [\ion{Ni}{1}]~3.12\um\ may contribute to the narrow peak on the blue side of the [\ion{Ca}{4}] emission.

[\ion{Ar}{2}]~6.98\um\ and [\ion{Ca}{2}]~$\lambda\lambda7291,~7324$ show similar line profiles (see \autoref{fig:IGE_IME_comp}), and the [\ion{Ca}{2}] line is helpful for interpreting the general shape of the low-resolution [\ion{Ar}{2}] line. \autoref{fig:IMEs_ionization} shows that the [\ion{Ar}{3}]~8.99\um\ line shape fits nicely with the [\ion{Ar}{2}] 6.98\um\ line if a blueshifted component is added. Thus, we fit [\ion{Ar}{2}]~6.98\um\ by summing a Gaussian with a forced contribution from the shape of the [\ion{Ar}{3}]. Since the [\ion{Ca}{2}] shares a similar shape to [\ion{Ar}{2}], with a Gaussian-like peak and a somewhat boxy red wing, we also fit it with the sum of a Gaussian and a forced contribution from the shape of the [\ion{Ar}{3}]. For [\ion{Ar}{2}] the additional Gaussian component has FWHM $=\,9300 \pm 1700$\kms and a kinematic offset $v_{\text{offset}}=\,-2300 \pm 100$\kms\ and for [\ion{Ca}{2}] the additional Gaussian component has FWHM $=\,4700 \pm 200$\kms and a kinematic offset $v_{\text{offset}}=\,-1500 \pm 100$\kms. Because the optical 0.73\um\ region is complex with contribution from many lines, including [\ion{Fe}{2}], [\ion{Ni}{2}], and [\ion{Ar}{3}], we fit all of these additional lines simultaneously with the [\ion{Ca}{2}] doublet and find their contribution to be relatively small. While we cannot definitively rule out other line profile shapes for [\ion{Ca}{2}] and [\ion{Ar}{2}], it is clear that they exhibit an additional blueshifted peak of emission not seen in the higher IME ionization states.

The shapes of these IME line profiles and their ionization stratification point to the presence of broad, asymmetric shells of higher ionization IME material, and a somewhat narrower, blueshifted region of lower ionization IME material.

\subsection{O and Ne}

In the center-bottom panel of \autoref{fig:all_fits}, we display our best-fit line profiles for [\ion{O}{1}] and [\ion{Ne}{2}]. The [\ion{O}{1}]~$\lambda\lambda6300,~6364$ feature is remarkably narrow (FWHM $= 2500 \pm 100$\kms), allowing us to resolve emission from both components of the doublet. We do not detect emission from either [\ion{O}{2}]~$\lambda$3727 or [\ion{O}{3}]~$\lambda\lambda$4959,~5007; however, this region of the spectrum is dominated by low-ionization-state IGEs, complicating the interpretation of the blended features. The [\ion{O}{1}] emission is best modeled by a single skewed Gaussian velocity component. Similar skewed (or ``sawtoothed") emission has been observed in the [\ion{Fe}{2}] and [\ion{Ca}{2}] features of other 03fg-like SNe~Ia \citep{Taubenberger2017, Siebert2023, Siebert2023b}. This asymmetry could potentially be explained by either an intrinsically asymmetric ejecta distribution or dust extinction from within the SN ejecta. 

If a large amount of dust were present in the ejecta, assuming it did not render the [\ion{O}{1}] emission invisible,  one would expect the redshifted emission to be more heavily extinguished than the blueshifted emission. This would lead to an emission profile whose total offset is preferentially blueshifted, appears steeper on the blueshifted side, and is more shallow-sloped on the redshifted side \citep{jerkstrand_spectra_2017}. There is some evidence that the [\ion{Ca}{2}] features of 03fg-like SNe are preferentially blueshifted \citep{Siebert2023b}; however, one would expect more dust attenuation at shorter wavelengths. Furthermore, the [\ion{O}{1}] in SN~2022pul is redshifted by $260 \pm 10$\kms, so the systemic velocity of the ejecta (and dust) would need to be redshifted if dust extinction were the cause of the asymmetric line profile. This could be the case; however, we do not observe evidence for such a systemic shift in the other nebular features, and therefore favor an asymmetric ejecta distribution. 

The [\ion{Ne}{2}] 12.81~$\mu$m emission has a unique profile compared to the other elements. There is a narrow peak close to $\sim 0$\kms, indicating that it is present in the central regions of the ejecta along with [\ion{O}{1}]. However, most of the [\ion{Ne}{2}] emission is likely not coincident with the [\ion{O}{1}] owing to its much larger ionization potential. The overall broad [\ion{Ne}{2}] line suggests that it forms at large velocities similar to the IMEs and IGEs. We model the broad component of the [\ion{Ne}{2}] line profile with a broad Gaussian on the red side (FWHM $= 10000 \pm 500$\kms), and an asymmetric slanted flat-top on the blue side (FWHM $=\,7200 \pm 600 $\kms, kinematic offset $v_{\text{offset}}=\,1400 \pm 200$\kms, inner-shell velocity $v_{\text{min}}=\,7000 \pm 500$\kms, and offset of the shell thickness $v_{\text{offset, shell}}=\,800 \pm 100$\kms). The one-sided asymmetric slant is tilted in the opposite direction as the tilt of the IMEs, and could arise from a half-shell or some type of missing bubble of [\ion{Ne}{2}] on the blueshifted side (see \autoref{fig:all_fits} for visualization of this profile). The overall shape of this profile does not appear as blueshifted as the other IMEs, and the differing locations of missing emission indicate that the distributions of Ne and the IMEs are distinctly different. 

The highly asymmetric [\ion{Ne}{2}] line shares some of the same projected velocities, but it cannot coexist with the IMEs and IGEs, indicating large-scale composition asymmetry. Additionally, the small-scale structure seen throughout the [\ion{Ne}{2}] profile may be caused by the presence of a large number of randomly distributed blobs \citep[e.g., see][Fig. 3]{jerkstrand_spectra_2017}; alternatively, it may be due to higher noise levels at the longest MIRI/LRS wavelengths. Other higher resolution optical and NIR emission lines, including the [\ion{O}{1}]~$\lambda\lambda6300,~6364$ feature, do not show convincing evidence for small structure in the line profiles that would suggest clumping on large scales (see \autoref{sec:model}). In \autoref{sec:conclusions}, we discuss the detection of [\ion{Ne}{2}] 12.81~$\mu$m in relation to predictions from WD--WD violent merger models.

\subsection{Geometric Interpretation: Distinct Ejecta Distributions}

The difference in line profiles suggests differences in the geometric distributions of material in the ejecta. In order to more easily compare between the bulk emission coming from each group of elements (IGEs: Fe, Co, Ni; IMEs: S, Ar, Ca; O, Ne), rather than from particular elements and ions, we sum all of the line profiles across our full wavelength range within these three groups in velocity space and calculate the bolometric luminosity along the line of sight in bins of 500\kms\ ($d\nu$ $L_{\text{bol}}$). Each velocity bin represents a slab of emission on the plane of the sky, giving a 2D projected spatial distribution of the elements. Shown in the bottom right panel of \autoref{fig:cartoon_distributions}, the summed profiles of the IGEs, IMEs, and O+Ne are close to those of the individual profiles, but they erase the small variations seen from profile to profile.

Using the geometric interpretations of the line-profile components that we adopt in our fits, we construct cartoons for the projected 2D distribution of our three element groups to aid visualization of the distinct regions (see \autoref{fig:cartoon_distributions}). Assuming spherical symmetry, Gaussian emission distributions produce Gaussian line profiles, and slanted flat-topped (or boxy) profiles result from asymmetric shells of emission. We assume spherical symmetry for the individual components of each profile (resulting in overall axial symmetry for the 2D projected distributions) and stick to combinations of these two basic line-profile shapes, so our cartoons are oversimplifications. The real ejecta are probably asymmetric and complex, potentially including bubbles, mushroom shapes, and clumps. \citet{Pakmor2012} show the distribution of elements in their WD violent merger hydrodynamical simulation in their Fig. 2. Interestingly, our cartoons bear some similarities to these physically realistic distributions.

We stress that these cartoons are not to be taken literally and that they are 2D projections. Overlap in line-of-sight velocity ($v_\text{LOS}$) between the element groups does not necessarily imply that they are physically co-located in 3D space. Specifically, all of our distributions show emission at $v_\text{LOS}=0$\kms; however, this is a projection effect because, from theoretical models, there should be very little to no IGEs or IMEs in the most central regions where the [\ion{O}{1}] emission is located.

\begin{figure*}
    \centering
    \includegraphics[width=\textwidth]{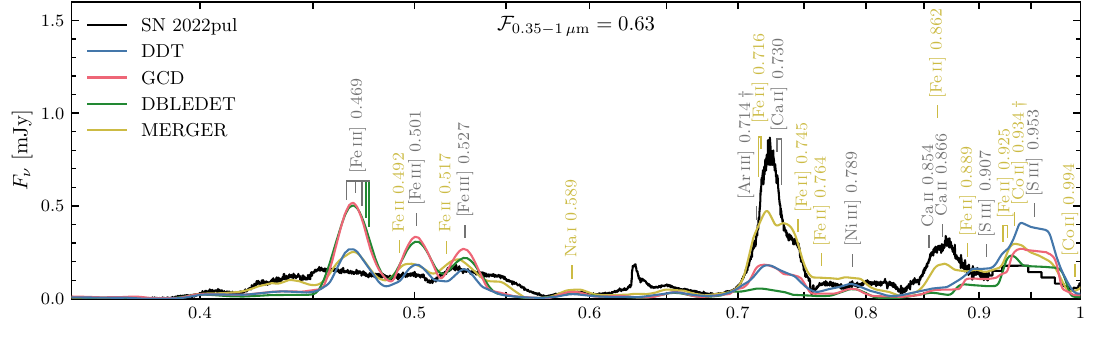}\vspace{-0.5cm}
    \includegraphics[width=\textwidth]{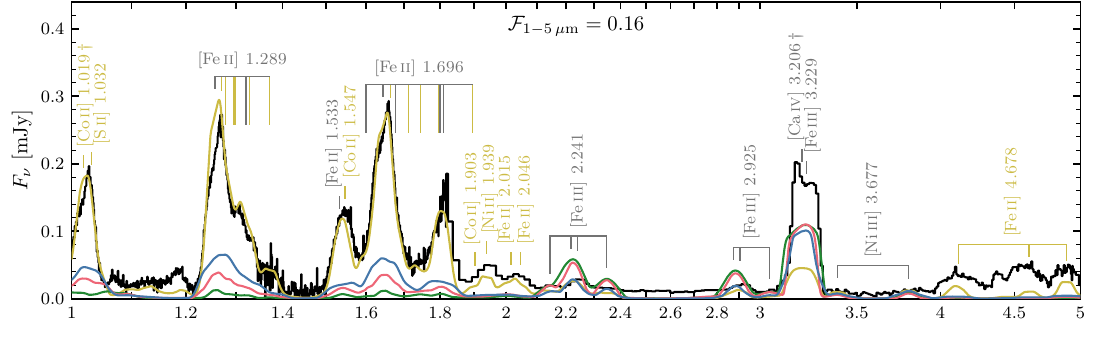}\vspace{-0.5cm}
    \includegraphics[width=\textwidth]{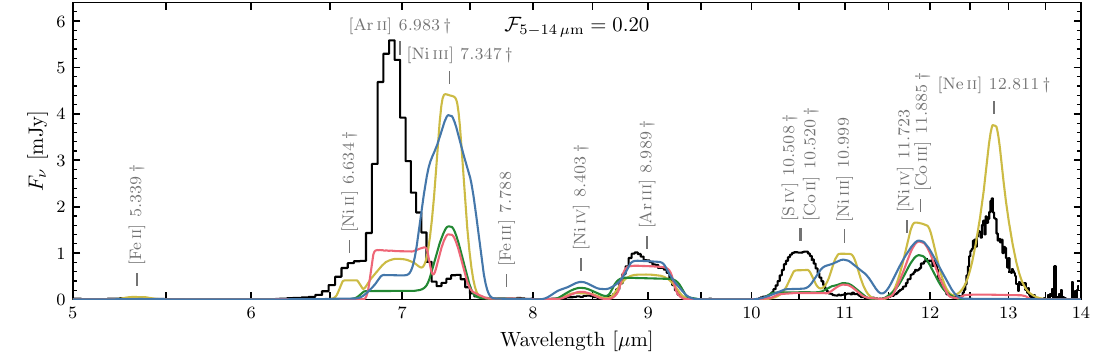}\vspace{-0.2cm}
   \caption{Spectra of our reference model set at 338 days post explosion compared to the dust and MIR-continuum-subtracted spectrum of SN 2022pul over the wavelength ranges (from top to bottom) optical (0.35--1\um), NIR (1--5\um), lower MIR (5--14\um), and upper MIR (14--28\um). The SN 2022pul spectrum has been corrected for redshift and MW extinction. The synthetic fluxes correspond to the assumed distance to SN 2022pul of 16 Mpc; they have not been rescaled or normalized in any way. The $F_{X-Y \text{$\mu$m}}$ label gives the fraction of the total optical to MIR flux (0.35--14\um) for SN~2022pul emitted within the wavelength range of each plot. We include line identifications based on their Sobolev equivalent width, as in \citet{Blondin2023} (see \autoref{tab:lineid}). Transitions connected to the ground state are marked with a “†” symbol. For consecutive lines within a multiplet (connected by a horizontal line), we give the $gf$-weighted mean wavelength of the transitions.} %Lines that only appear in one model class are labeled with the corresponding color.}
    \label{fig:all_models}
\end{figure*}

\begin{figure*}
    \centering
    \includegraphics[width=1.0\textwidth]{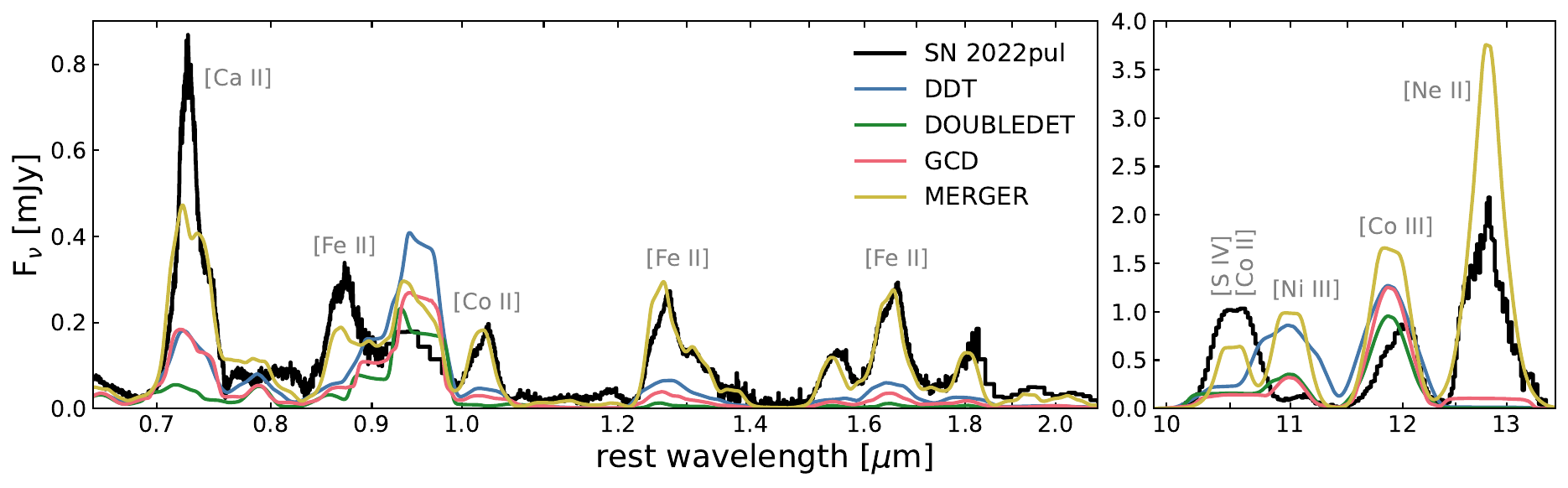}
    \caption{Similar to \autoref{fig:all_models} but highlighting the 0.7--2\um\ and 10--14\um\ regions where the models diverge substantially. The MERGER model matches the observed spectrum of SN~2022pul best, notably producing stronger [\ion{Ca}{2}] near 0.73\um, capturing the NIR [\ion{Fe}{2}] lines extremely well, and predicting the strong observed [\ion{Ne}{2}] 12.81\um\ line that is not seen in the other models. However, there are also deficiencies to the MERGER model, such as the strength of the Ni lines being overpredicted (e.g., [\ion{Ni}{3}]~11.00\um).}
    \label{fig:full_spec_model}
\end{figure*}

\subsection{Nickel-56 Mass} % avoid special characters in section headings
At the late phase of SN~2022pul (338~d post explosion), the $^{56}$Ni (half-life of $\sim 6$~d) has long since decayed away and the energy deposition that continues to power the spectral emission lines comes from the radioactive decay of $^{56}$Co (half-life of $\sim 77$~d). Locally-deposited gamma-rays and positrons from the decay can be absorbed by the IGEs and some IMEs in the IGE-rich regions, while the IMEs and unburned elements outside this region can absorb nonlocally deposited gamma-rays and secondary, low-energy photons. Thus, to estimate the initial mass of $^{56}$Ni synthesized in the explosion, we account for the total flux by integrating the spectrum ($F = \int F_\nu\,d\nu$) in the range 0.4--14\um. We then estimate the luminosity from the emission lines by correcting the flux to a distance of 16 Mpc (see Paper I, \citealt{Siebert2023b}, for a discussion of distance to SN~2022pul). In the low-density nebular phase, however, most of the energy escapes. To account for this, we divide by the fraction of the total decay energy that is absorbed by the ejecta from the MERGER model \citep[][and described further in \autoref{sec:model}]{Blondin2023}, which has a value $E_\text{absorbed}/E_\text{total}=0.055$. 

Using the uncertainty in the distance (16 $\pm$ 2~Mpc) as the dominant source of uncertainty in our calculation, we estimate a bolometric luminosity (i.e., deposited $^{56}$Co-decay power) at 338~d post explosion of $4.6 \pm 1.2\times 10^{41}$~erg~s$^{-1}$, which corresponds to a mass of synthesized $^{56}$Ni of $0.66 \pm 0.17$\msun \citep[see equations from][Section 5.4]{Nadyozhin1994,Branch2017}. This estimate may also suffer from uncertainties in reddening, absolute-flux calibration, gamma-ray escape fraction, and assumption of spherical symmetry, but these uncertainties should be smaller than the 25\% uncertainty introduced by the distance. Our estimated $^{56}$Ni mass is in rough agreement with the 0.6 \msun of $^{56}$Ni produced in the violent merger simulation by \citet{Pakmor2012} and the calculated $^{56}$Ni masses for SN~2011fe, the canonical normal SN~Ia \citep[][and references therein]{Bora2022}.

\section{WD--WD Violent Merger Model \label{sec:model}}

\begin{figure*}
    \centering
    \includegraphics[width=\textwidth]{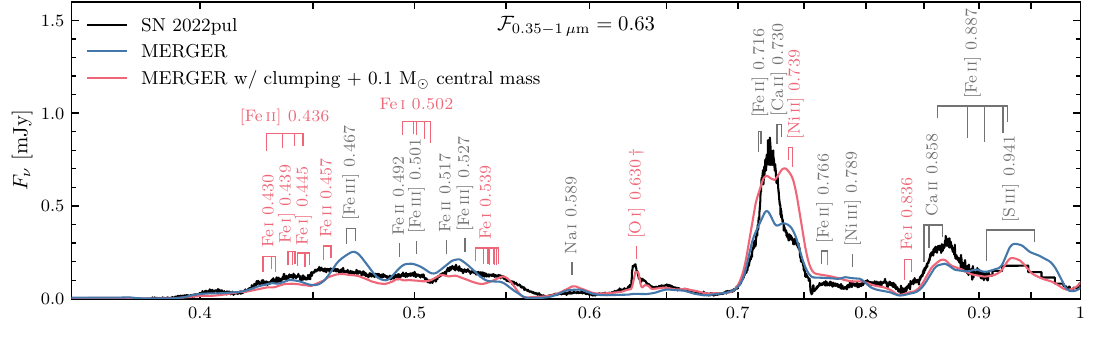}\vspace{-0.5cm}
    \includegraphics[width=\textwidth]{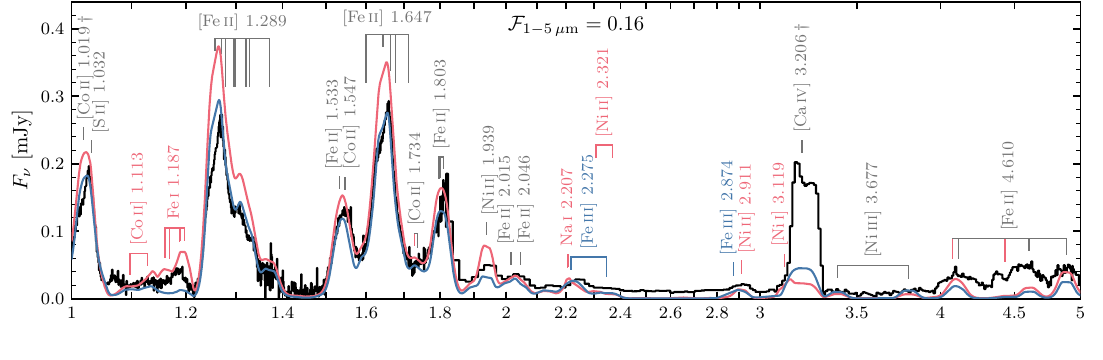}\vspace{-0.5cm}
    \includegraphics[width=\textwidth]{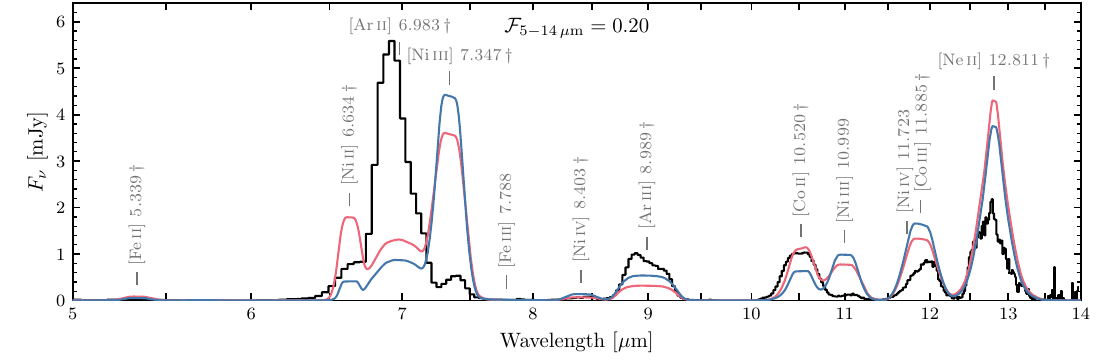}\vspace{-0.2cm}
    \caption{Similar to \autoref{fig:all_models} but for our reference (blue) and modified (red) merger models.}
    \label{fig:model_zoomins}
\end{figure*}

\begin{figure*}
    \centering
    \includegraphics[width=0.85\textwidth]{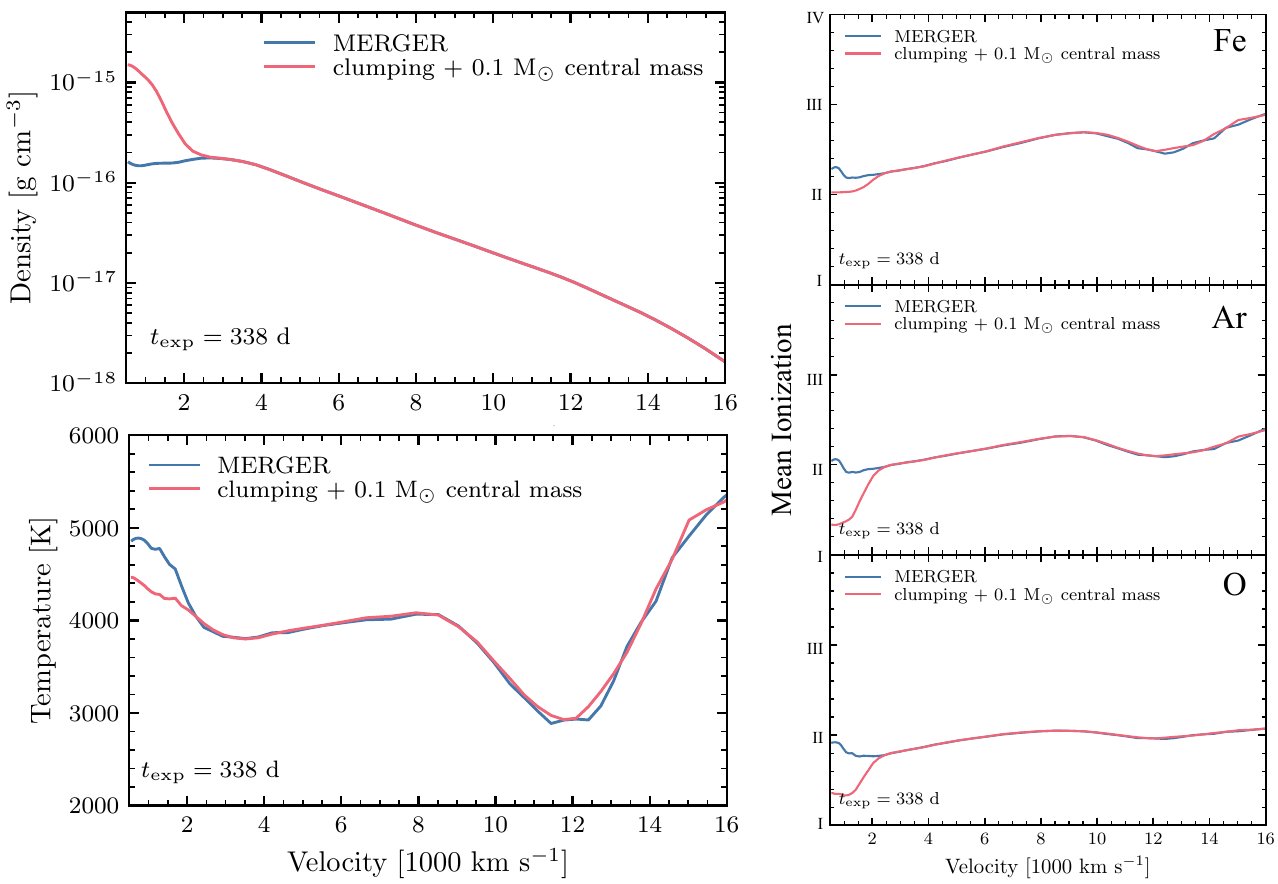}
    \caption{\textit{Top left:} Density profiles at 338~d for the MERGER model and our modified merger model with clumping and 0.1\msun\ of mass added to the central ($\lesssim 2000$\kms) region. \textit{Bottom left:} Temperature profiles. \textit{Top right:} Mean Fe ionization profiles. The mean ionization is defined as $\Sigma_i in^{i+} / \Sigma_i n^{i+}$, where $n^{i+}$ is the number density of ionization stage $i$ for Fe, such that a mean ionization of $\sim 2$ indicates that Fe$^{2+}$ is the dominant stage. \textit{Middle right:} Mean Ni ionization profiles. \textit{Bottom right:} Mean O ionization profiles.}
    \label{fig:ejecta_prop}
\end{figure*}

\cite{Blondin2023} presented nebular-phase spectra from four different SN~Ia explosion models at 270~d post-explosion (a \mch\ delayed-detonation model [DDT; \citealt{Seitenzahl2013}], a \mch\ pulsationally-assisted gravitationally-confined detonation model [GCD; \citealt{Lach2022}], a sub-\mch\ double-detonation [DBLEDET; \citealt{Gronow2021}], and a violent merger of two sub-\mch\ WDs (0.9$\,+\,$1.1\,\msun) [MERGER; \citealt{Pakmor2012}]), which they compared to a spectrum of SN~2021aefx covering the wavelength range 0.3--14~$\mu$m from \cite{Kwok2023}. These model spectra are generated assuming spherical symmetry, so the line profile shapes do not reflect ejecta asymmetries that might be seen in two-dimensional models. Only the MERGER model displayed a strong, centrally-peaked line due to [\ion{Ne}{2}]\,12.81\um, resulting from the presence of Ne in the innermost ejecta layers of this model. None of the other models considered by \cite{Blondin2023} presented such a strong [\ion{Ne}{2}] feature, and only the GCD model displayed a weak flat-topped line, whose intensity at line center was a factor of $\sim3$ smaller than observed in SN~2021aefx. Only by artificially setting a minimum Ne mass fraction of $10^{-2}$ in the inner ejecta of the GCD model were they able to match the peak intensity and profile shape of the potential weak [\ion{Ne}{2}] line in SN~2021aefx. Future analysis of subsequent observations of SN~2021aefx should clarify whether [\ion{Ne}{2}] 12.81\um\ is actually present in SN~2021aefx (Ashall et al., in prep.).

The violent merger model naturally predicts the presence of Ne in the inner ejecta layers, which is associated with the delayed and incomplete burning of the lower-mass secondary WD. Moreover, the higher density in layers $\lesssim 10,000$\kms\ enhances the recombination rate there, which in turn results in a lower mean ionization state compared to the other explosion models. For comparing to the nebular spectrum of SN~2022pul, we recomputed the four models studied by \cite{Blondin2023} at a later time of 338~d post-explosion, using the same numerical setup. The full optical through MIR comparison of these models to the observed spectrum is displayed in \autoref{fig:all_models} and highlights of the results are shown in \autoref{fig:full_spec_model}.

Given the limited range of models from very different origins, we compare qualitatively rather than quantitatively. Of the four models, we favor the MERGER model for several reasons: (1) it predicts a lower overall ionization state that is in good agreement with the NIR [\ion{Fe}{2}] lines; (2) it predicts rounded IME emission profiles, indicating that the IMEs are centrally located, whereas the other models predict very flat profiles for all ionization species, indicating that the IMEs are absent in the center; and (3) it is the only model consistent with the strong, centrally peaked [\ion{Ne}{2}]\,12.81\um\ line.

The lower ionization results in a better match to the numerous [\ion{Fe}{2}] features in the 1--2~$\mu$m range, yet the optical [\ion{Fe}{3}] complex around 0.5\um\ is still too strong compared to SN~2022pul, as is the [\ion{Ni}{3}]\,7.35\um\ line. This suggests the ionization in SN~2022pul is even lower than predicted in the MERGER model. We consider clumping to be a viable mechanism to further reduce the ionization state in this model. Using the same approach as \cite{Blondin2023}, we recomputed a clumped version of the MERGER model with a volume-filling factor $f=0.5$ (i.e., resulting in a factor of two increase in density of the clumps). Clumpy ejecta can leave an impact on emission lines (e.g., nebular spectra of the Type Ib SN 1985F and of the Type IIb SN 1993J; \citealt{Filippenko1989, Filippenko1994, Houck1996}), but these would need to be large clumps with a large density contrast (i.e., dense shells in 1D or large clumps in 3D), whose physical origin in SN Ia ejecta is unclear. In our modeling approach the clumps are significantly smaller than the photon mean free path, so we see a global impact on the ionization state (through enhanced recombination) but no small structure in the line profiles.

As seen in \autoref{fig:model_zoomins}, this clumped model is able to better match the flux level in the [\ion{Fe}{3}]-dominated complex, while maintaining a satisfactory match out to $\sim 3$\um. The lower ionization also results in a stronger [\ion{Ar}{2}]\,6.98\um\ line and a weaker [\ion{Ni}{3}]\,7.35\um\ line. However, it also results in an even weaker [\ion{Ca}{4}]\,3.12~$\mu$m line compared to SN~2022pul. The same is true of [\ion{Ar}{3}]\,8.99\um. Moreover, the [\ion{Ni}{2}]\,6.63\um\ and [\ion{Ni}{3}]\,7.35 and 11.00\um\ lines remain too strong. This suggests that the clumping is not uniform, affecting some regions more strongly than others.

The mismatch in lines from specific elements over several ionization stages suggests that abundance also plays a role. Here, a lower abundance of stable Ni and a higher abundance of Ar/Ca could help resolve some of these discrepancies. In particular, a lower abundance of stable Ni ($\sim0.03$\,\msun\ in this MERGER model) seems warranted to reproduce the lines of \ion{Ni}{2} and \ion{Ni}{3} in the observed spectrum. \citet{Pakmor2013} show in their helium-ignited violent merger of a system of 0.9$\,+\,$0.76\,\msun\ WDs that the central density and the Ar/Ca abundance is higher in the innermost layers, while the stable Ni abundance is lower and more centrally concentrated than the merger from \citet{Pakmor2012}, which the MERGER model is based on. Additionally, the amount of stable Ni in the merger model from \citet{Pakmor2012} scales with the metallicity of the progenitor system ($Z=Z_\odot$ in the model), so lower stable Ni abundance may imply the progenitor system had subsolar metallicity. There may be a variety of inner-ejecta properties that could be produced through violent merger scenarios that would better match the observations of SN~2022pul.

Despite the lower ionization of the clumped MERGER model, the [\ion{O}{1}]~$\lambda\lambda6300,~6364$ emission remains very weak compared with SN~2022pul. This is not related to an abundance issue, since oxygen completely dominates the ejecta layers below $\sim 2000$\,\kms\ in the MERGER model (mass fraction $\gtrsim 0.5$). Radial compression does not change the optical depth, so the mass of inner material must be increased to boost the amount of gamma-ray energy intercepted by the inner, O-rich ejecta layers. We were able to match the intensity and width of the narrow [\ion{O}{1}] feature seen in SN~2022pul by artificially increasing the density below $\sim 2000$\kms. In practice we followed the procedure from \citet{Dessart2020} and added a Gaussian component to the original density profile centered on the inner boundary of our spatial grid ($v_\mathrm{min}\approx 580$\kms) with a width $\sigma\approx 600$\kms, resulting in an additional 0.1\,\msun\ in these layers ($\sim 50$\% of which is O). The original mass of O (0.53\msun) was all located in the inner ejecta and the total ejecta mass in the model is increased through this modification. The fractional O abundance in the innermost region is not particularly important, as long as [\ion{O}{1}]~$\lambda\lambda6300,~6364$ remains the dominant coolant.

As seen from \autoref{fig:model_zoomins}, the narrow [\ion{O}{1}] feature is well reproduced, while the rest of the spectrum remains largely unchanged compared to the original model. The physical origin of such an increase in density in the inner ejecta is likely the result of a higher-mass secondary WD (e.g., \citealt{Pakmor2013}). Other contributing effects could be how completely the secondary WD burned, a different C/O fraction of the initial secondary WD, and compression of the innermost layers through the $^{56}$Ni bubble effect \citep[e.g.,][]{Wang2005,Dessart2021}. The changes to the ejecta properties (density, temperature, mean ionization) in our modified merger model where clumping and 0.1~$M_\odot$ central mass have been introduced are shown in \autoref{fig:ejecta_prop}, and a detailed list of important emission lines in our modified merger model is given in \autoref{tab:lineid}.

While additional fine tuning of the abundances and densities of particular elements could help improve the match of any of the models, only the MERGER model can be modified to create the emergence of the narrow, central [\ion{O}{1}] feature. This is because the other models predict essentially no O or Ne in the low-velocity central regions, while the MERGER model is actually dominated by O in the center (mass fraction $\sim$0.5). Fine tuning would not produce the emergence of the central O and Ne lines in the other models because they are not dominant coolants in this region, and adding a significant central mass of O and Ne to the other models would be unphysical.

\input{lineid_merger_clumped_7}

\section{Discussion \& Conclusions \label{sec:conclusions}}

As discussed in Paper I \citep{Siebert2023b}, the detection of [\ion{O}{1}] emission in WD SNe is exceedingly rare. SN~2022pul is now the third 03fg-like SN~Ia to exhibit [\ion{O}{1}], along with SN~2012dn \citep{Taubenberger2019} and SN~2021zny \citep{Dimitriadis2023}. The velocity distribution of [\ion{O}{1}] varies greatly among these SNe, while its offset is consistent with originating from the central regions of the ejecta in each case. Additionally, SN~2022pul exhibits [\ion{Ne}{2}] peaked in the central regions and extending farther out as well. A viable explosion model should be able to explain the presence and diversity of oxygen and neon. Both turbulent pure deflagrations \citep{Kozma2005, Fink2014} and violent WD mergers \citep{Pakmor2012, Kromer2013, Pakmor2013} naturally predict neon and unburned oxygen present at low velocities. A violent WD--WD merger provides an explanation for the large-scale asymmetries in the nebular emission features described in \autoref{sec:line_profiles}, though a turbulent deflagration could also produce asymmetries. However, a pure deflagration would produce a low-luminosity explosion with different early-time spectroscopic properties, narrower nebular line profiles owing to the lower explosion energy, and weaker emission from IMEs. To check this, we ran the N100def model from \citet{Fink2014} with CMFGEN to 338~d post explosion (not shown here) and indeed found that the MERGER model is a significantly better match to SN~2022pul for the above reasons. We therefore favor a WD merger over a pure deflagration for SN~2022pul.

The large parameter space in WD mergers could help explain the diversity of 03fg-like SNe. The variation among objects may be explained by different WD mass ratios, by the specific explosion mechanism of one or both WDs, and by observations along different lines of sight. Recent studies of 03fg-like SNe Ia have favored WD merger models occurring within a dense C/O-rich CSM \citep{Dimitriadis22, Srivastav2023, Dimitriadis2023, Siebert2023} due to rapidly evolving light-curve bumps, low-ionization nebular spectra sometimes with [\ion{O}{1}], asymmetric and blueshifted nebular emission lines, and a larger fraction of flux in the NIR at late times than normal SNe Ia. Our observations of SN~2022pul enhance the evidence for this model for 03fg-like SNe Ia through clear dust detection, unusual light-curve rise shape, asymmetric line profiles, strong IMEs, and central O and Ne.

In the case of SN~2022pul, we also observe distinct ejecta components with different composition offset in their bulk velocities. The secondary WD may have been partially, but not completely, disrupted when the primary WD detonated. When burning front passes over the lower density secondary WD material, it too can burn, synthesizing O, Ne, and the IMEs (whereas the higher density primary WD would produce the IGEs) \citep{Pakmor2017}. The degree of ejecta asymmetry will be smaller if the secondary is more compact when the primary is burned because the ashes of the primary will be less hindered by a less disrupted secondary \citep{Pakmor2017}.

In SN~2022pul, we observe central O and Ne, very strong emission from IMEs, and a high degree of asymmetry in the nebular line features. Furthermore, our modified merger model demonstrates that additional, concentrated central mass (0.1~M$_\odot$ at $\lesssim 2000$\kms) is needed to reproduce the [\ion{O}{1}]~$\lambda\lambda6300,~6364$ emission. This very centrally peaked and asymmetric [\ion{O}{1}] emission is reminiscent of expectations of stripped material from a companion \citep{Marietta2000,Botyanszki2018}, and may indicate that the secondary WD left a significant amount of mass in the innermost region. 

The central, broad Ne and strong IMEs suggest that the secondary WD material was at least partially burned (the O may be a combination of burned C and unburned O, while unburned Ne is not expected for the lower-mass secondary). A higher mass, and therefore less disrupted secondary WD, could produce the strong IMEs and the centrally concentrated material. However, the high degree of asymmetry would suggest a more disrupted secondary, which perhaps could have left behind centrally concentrated material if it was only partially burned. This may be due in part to a viewing-angle effect where the ejecta appear more symmetric in the vertical direction, and more asymmetric in the plane of rotation \citep{Bulla2016, Pakmor2017}. Previous measurements of 03fg-like SNe Ia have not shown strong continuum polarization \citep{Tanaka2010, Cikota2019}. More polarization data and construction of nebular spectra along different viewing angles from violent merger hydrodynamical simulations would help resolve this and improve our understanding of the 3D ejecta distributions giving rise to the asymmetric line profiles.

Another key part of understanding this progenitor scenario is its connection to the inferred presence of a dense C/O-rich CSM from the shape of the light-curve rise \citep{Siebert2023b}. This requires considering scenarios where the explosion occurs either pre-merger during a dynamical interaction phase (commonly known as a ``perimerger"), or scenarios where explosion occurs post-merger after the complete disruption of the secondary WD \citep{Pakmor2010, vanKerkwijk2010, Pakmor2012,Moll2013, Guillochon2010, Raskin2013}.

In both cases, simulations of the ejecta are highly asymmetric and the synthetic light curves cover a large range in luminosity depending on viewing angle (with the highest luminosity observed along the equator). This light-curve line-of-sight dependence might also vary over time because at early phases the interaction dominates over decay power. One beneficial consequence of the post-merger scenario is that the complete disruption of the secondary WD leads to the formation of a massive C/O-rich accretion disk. The early-time light curve of SN~2022pul is consistent with ejecta interaction with this kind of CSM \citep{Noebauer2016,Siebert2023b}. More work is needed to understand if a violent merger is capable of producing enough CSM at the required distance to explain this kind of interaction signature.  

The large amount of CSM (0.6\msun) adopted in the \citet{Noebauer2016} interaction model is difficult to reconcile with the central oxygen and neon seen in SN~2022pul. If this much material from the secondary WD was pulled off into an exterior spherical shell, it seems unlikely that the secondary would have enough leftover mass to create the high-density central O giving rise to the [\ion{O}{1}] doublet and the strong emission from the IMEs. If the CSM is responsible for damping the kinetic energy of the explosion, leading to the boosted luminosity and low-velocity \ion{C}{2} at early times, the CSM must cover a substantial fraction of the exploding system \citep{maeda23}. \citet{howell_type_2006} suggest that the extra binding energy associated with the additional mass may be enough to cause sufficiently slow ejecta velocity. However, \citet{Raskin2013} find that the amount of CSM stripped from the secondary WD in their models is small, less than 0.005\msun. If the dust emission arises from CSM, assuming a dust-to-gas ratio of 0.01, the dust mass for SN~2022pul given by \citet{Johansson2023} implies a CSM mass of 0.05--0.11\msun, depending on composition. Further exploration is needed to address whether these lower CSM masses (or perhaps a different composition) could produce the interaction signature we observe.

\citet{Hsiao2020} present evidence supporting the ``core-degenerate'' model for LSQ14fmg, where a near-\mch\ WD interacts with the degenerate core of an AGB star within a common envelope stripped of its H and He through a superwind phase prior to explosion. We disfavor this scenario to explain SN~2022pul because the explosion is expected to be a spherical explosion of a single object after secular accretion. It therefore does not explain the observed central location of O and Ne or the large-scale ejecta asymmetries. It is tempting to speculate whether a modified version of this model, in which the AGB degenerate core material also undergoes thermonuclear fusion, could be developed for SN~2022pul.

Perhaps a triple system could be invoked \citep{kushnir_head-on_2013, Grishin2022}, where one WD or star is disrupted to become the surrounding CSM and precipitates the violent merger of the remaining two WDs. \citet{Rajamuthukumar2023} found that triple systems contribute significantly to the overall SN~Ia rate. Again, more detailed modeling over a wide range of possible parameters and scenarios involving WD--WD violent mergers is needed to better understand the system giving rise to this unique SN.

We summarize our main observations and discussion points below.
\begin{itemize}
    \item The presence of O and Ne in the central ejecta are hallmarks of the WD violent merger SN~Ia model.
    \item Asymmetric nebular emission-line profiles show distinct distributions of IGEs, IMEs, and O and Ne in the ejecta that can be naturally explained by the violent merger model.
    \item A pure deflagration model is disfavored as it would produce lower luminosities, narrower nebular line widths, and weaker emission from IMEs than observed in SN~2022pul.
    \item Especially strong emission from IMEs could come from burning of a disrupted, lower-density secondary WD.
    \item Our spectrum of SN~2022pul is quite well matched by the MERGER model nebular spectrum computed by \citet{Blondin2023} from the violent WD--WD merger model of \citet{Pakmor2012}. The agreement between the observed and model spectra is further improved by the addition of clumping and 0.1\msun\ to the  central mass.
    \item It is still somewhat unclear how the violent merger model can produce both the C/O-rich CSM described by Paper I \citep{Siebert2023b} and Paper III \citep{Johansson2023}, and the central O and Ne seen in the spectra.
    \item Further work and more detailed models are needed to constrain the masses of the WDs involved in the merger.
\end{itemize}

Our observations of SN~2022pul are a prime example of the importance of gathering spectra of WD SNe across the optical, NIR, and MIR wavelengths. The MIR observation of [\ion{Ne}{2}] is complementary to the observation of [\ion{O}{1}] in the optical; the optical [\ion{Ca}{2}], NIR [\ion{Ca}{4}], and MIR [\ion{Ar}{2}], [\ion{Ar}{3}], and [\ion{S}{4}] mutually help interpret the IME distribution and ionization stratification; and the NIR [\ion{Fe}{2}] and MIR [\ion{Co}{3}] lines establish the distribution of the IGEs. Furthermore, the higher resolution ground-based optical and NIR observations were essential to interpret the lower resolution \textit{JWST} observations, and isolated MIR spectral lines were key to unraveling the complex line blending in the optical and NIR. 

Future observations of 03fg-like SNe~Ia would benefit from \textit{JWST} spectroscopy, especially in the higher resolution instrument modes which reach to even longer wavelengths and reveal additional lines. Complementary spectropolarimetry would give insight into line-profile asymmetries and constrain aspect angles \citep{DerKacy2023}. If this class of objects indeed arises from WD mergers, the observations should show a large degree of diversity in their asymmetric line profiles, and \textit{JWST} observations could additionally establish a firmer connection to the dust surrounding or formed in these objects.\\
\vspace{1cm}

% \begin{acknowledgements}

This work is based on observations made with the NASA/ESA/CSA \textit{JWST} as part of program \#02072. We thank Shelly Meyett for her consistently excellent work scheduling the \textit{JWST} observations, Sarah Kendrew for assistance with the MIRI observations, and Glenn Wahlgren for help with the NIRSpec observations. We also thank the anonymous referee whose suggestions significantly improved the paper.
The data were obtained from the Mikulski Archive for Space Telescopes at the Space Telescope Science Institute (STScI), which is operated by the Association of Universities for Research in Astronomy (AURA), Inc., under National Aeronautics and Space Administration (NASA) contract NAS 5-03127 for \textit{JWST}. Support for this program at Rutgers University was provided by NASA through grant JWST-GO-02072.001. 

The SALT observations of SN~2022pul were obtained with Rutgers University program 2022-1-MLT-004 (PI S.~W.~Jha). We are grateful to SALT Astronomer Rosalind Skelton for taking these data.
This work makes use of data from the Las Cumbres Observatory global network of telescopes.  The LCO group is supported by NSF grants AST-1911151 and AST-1911225. This work also makes use of data gathered with the 6.5 meter Magellan telescopes at Las Campanas Observatory, Chile.

M.R.S. is supported by an STScI Postdoctoral Fellowship. G.D. acknowledges H2020 European Research Council grant \#758638. L.A.K. acknowledges support by NASA FINESST fellowship 80NSSC22K1599. C.L. is supported by an NSF Graduate Research Fellowship under grant \# DGE-2233066.
The UCSC team is supported in part by NASA grant NNG-17PX03C, National Science Foundation (NSF) grant AST-1815935, the Gordon and Betty Moore Foundation, the Heising-Simons Foundation, and a fellowship from the David and Lucile Packard Foundation to R.J.F. The work of A.V.F.'s supernova group at UC Berkeley is generously supported by the Christopher R. Redlich Fund, 
Gary and Cynthia Bengier, Clark and Sharon Winslow, Alan Eustace and Kathy Kwan, William Draper, Timothy and Melissa Draper, Briggs and Kathleen Wood, and Sanford Robertson (W.Z. is a Bengier-Winslow-Eustace Specialist in Astronomy, T.G.B. is a Draper-Wood-Robertson Specialist in Astronomy, Y.Y. was a Bengier-Winslow-Robertson Fellow in Astronomy), and many other donors. 

S.B. acknowledges support from the Alexander von Humboldt Foundation and from the ``Programme National de Physique Stellaire'' (PNPS) of CNRS/INSU cofunded by CEA and CNES.
D.J.H. receives support through NASA astrophysical theory grant 80NSSC20K0524.
J.V. and T.S. are supported by the NKFIH-OTKA grants K-142534 and FK-134432 of the Hungarian National Research, Development and Innovation (NRDI) Office, respectively.
B.B. and T.S. are supported by the \'UNKP-22-4 and \'UNKP-22-5 New National Excellence Programs of the Ministry for Culture and Innovation from the source of the NRDI Fund, Hungary. T.S. is also supported by the J\'anos Bolyai Research Scholarship of the Hungarian Academy of Sciences.
The research of J.C.W. and J.V. is supported by NSF grant AST-1813825.

A.F. acknowledges support by the European Research Council (ERC) under the European Union’s Horizon 2020 research and innovation program (ERC Advanced Grant KILONOVA \#885281).
M.D., K.M., and J.H.T. acknowledge support from EU H2020 ERC grant \#758638.
Research by Y.D. and S.V. is supported by NSF grant AST-2008108
Time-domain research by D.J.S. and the University of Arizona team is supported by NSF grants AST-1821987, 1813466, 1908972, and 2108032, and by the Heising-Simons Foundation under grant \#2020--1864. 
K.M. acknowledges support from the Japan Society for the Promotion of Science (JSPS) KAKENHI grant JP20H00174, and the JSPS Open Partnership Bilateral Joint Research Project (JPJSBP120209937). 
L.G. acknowledges financial support from the Spanish Ministerio de Ciencia e Innovaci\'on (MCIN), the Agencia Estatal de Investigaci\'on (AEI) 10.13039/501100011033, and the European Social Fund (ESF) ``Investing in your future'' under the 2019 Ram\'on y Cajal program RYC2019-027683-I and the PID2020-115253GA-I00 HOSTFLOWS project, from Centro Superior de Investigaciones Cient\'ificas (CSIC) under the PIE project 20215AT016, and the program Unidad de Excelencia Mar\'ia de Maeztu CEX2020-001058-M.

J.P.H. acknowledges support from the George A.\ and Margaret M.\ Downsbrough bequest.
The Aarhus supernova group is funded by a project 2 grant (10.46540/2032-00022B) from the Independent Research Fund Denmark (IRFD), and by a grant from Aarhus University Research Fond (AUFF-E-2023-9-28). 
This publication was made possible through the support of an LSSTC Catalyst Fellowship to K.A.B., funded through grant 62192 from the John Templeton Foundation to the LSST Corporation. The opinions expressed in this publication are those of the authors and do not necessarily reflect the views of LSSTC or the John Templeton Foundation.

Some of the data presented herein were obtained at the W.~M. Keck Observatory, which is operated as a scientific partnership among the California Institute of Technology, the University of California, and NASA. The Observatory was made possible by the generous financial support of the W.~M. Keck Foundation. The authors wish to recognize and acknowledge the very significant cultural role and reverence that the summit of Maunakea has always had within the indigenous Hawaiian community. We are most fortunate to have the opportunity to conduct observations from this mountain.
A major upgrade of the Kast spectrograph on the Shane 3~m telescope at Lick Observatory, led by Brad Holden, was made possible through generous gifts from the Heising-Simons Foundation, William and Marina Kast, and the University of California Observatories. Research at Lick Observatory is partially supported by a generous gift from Google.
We thank the Subaru staff for the data taken by the Subaru Telescope (S23A-023).

\facilities{AAVSO, ANU (WiFeS), ASAS-SN, ATLAS, GTC (OSIRIS), JWST (NIRSpec/MIRI), Keck:I (LRIS), Keck:II (NIRES), Keck:II (DEIMOS), LCO/GSP, Magellan (IMACS), MMT (Binospec), SALT (RSS), Shane (Kast), SOAR (Goodman), Subaru (FOCAS), UH (SNIFS), ZTF}

\software{Astropy \citep{astropycollaboration_astropy:_2013, astropycollaboration_astropy_2018, AstropyCollaboration2022}, 
Matplotlib \citep{hunter_matplotlib:_2007}, 
NumPy \citep{oliphant_guide_2006}, PyRAF \citep{Pyraf}, PySALT \citep{PySALT}, dust extinction \citep{karl_gordon_2022_6397654}, jdaviz \citep{jdadf_developers_2022_7255461}, jwst \citep{Bushouse_JWST_Calibration_Pipeline_2022}, UltraNest \citep{Buchner2021, johannes_buchner_2022_7053560}, {\tt YSE-PZ} \citep{CoulterZenodo, CoulterYSEPZ}, CMFGEN \citep{Hillier2012}
}

\clearpage

\normalsize
\bibliography{zotero_abbrev}

\end{document}

%% file: affiliations.tex
\newcommand{\LCO}{\affiliation{Las Cumbres Observatory, 6740 Cortona Drive, Suite 102, Goleta, CA 93117-5575, USA}}
\newcommand{\UCSB}{\affiliation{Department of Physics, University of California, Santa Barbara, CA 93106-9530, USA}}

\newcommand{\UCD}{\affiliation{Department of Physics and Astronomy, University of California, Davis, 1 Shields Avenue, Davis, CA 95616-5270, USA}}

\newcommand{\OKC}{\affiliation{Oskar Klein Centre, Department of Physics, Stockholm University, Albanova University Center, SE-106 91, Stockholm, Sweden}}

\newcommand{\UCB}{\affiliation{Department of Astronomy, University of California, Berkeley, CA 94720-3411, USA}}

\newcommand{\STScI}{\affiliation{Space Telescope Science Institute, 3700 San Martin Drive, Baltimore, MD 21218-2410, USA}}

\newcommand{\UA}{\affiliation{Steward Observatory, University of Arizona, 933 North Cherry Avenue, Tucson, AZ 85721-0065, USA}}

\newcommand{\Carnegie}{\affiliation{Observatories of the Carnegie Institute for Science, 813 Santa Barbara Street, Pasadena, CA 91101-1232, USA}}

\newcommand{\IfA}{\affiliation{Institute for Astronomy, University of Hawai`i, 2680 Woodlawn Drive, Honolulu, HI 96822-1839, USA}}
\newcommand{\UCSC}{\affiliation{Department of Astronomy and Astrophysics, University of California, Santa Cruz, CA 95064-1077, USA}}

\newcommand{\Princeton}{\affiliation{Department of Astrophysical Sciences, Princeton University, 4 Ivy Lane, Princeton, NJ 08540-7219, USA}}

\newcommand{\JHU}{\affiliation{Department of Physics and Astronomy, The Johns Hopkins University, 3400 North Charles Street, Baltimore, MD 21218, USA}}

\newcommand{\GeminiNorth}{\affiliation{Gemini Observatory/NSF's NOIRLab, 670 North A`ohoku Place, Hilo, HI 96720-2700, USA}}

\newcommand{\Rutgers}{\affiliation{Department of Physics and Astronomy, Rutgers, the State University of New Jersey,\\136 Frelinghuysen Road, Piscataway, NJ 08854-8019, USA}}
\newcommand{\FSU}{\affiliation{Department of Physics, Florida State University, 77 Chieftan Way, Tallahassee, FL 32306-4350, USA}}
\newcommand{\Melbourne}{\affiliation{School of Physics, The University of Melbourne, Parkville, VIC 3010, Australia}}
\newcommand{\ASTROthreeD}{\affiliation{ARC Centre of Excellence for All Sky Astrophysics in 3 Dimensions (ASTRO 3D)}}

\newcommand{\TAMU}{\affiliation{Department of Physics and Astronomy, Texas A\&M University, 4242 TAMU, College Station, TX 77843, USA}}
\newcommand{\Mitchell}{\affiliation{George P.\ and Cynthia Woods Mitchell Institute for Fundamental Physics \& Astronomy, College Station, TX 77843, USA}}

\newcommand{\ICE}{\affiliation{Institute of Space Sciences (ICE, CSIC), Campus UAB, Carrer de Can Magrans, s/n, E-08193 Barcelona, Spain}}
\newcommand{\IEEC}{\affiliation{Institut d’Estudis Espacials de Catalunya (IEEC), E-08034 Barcelona, Spain}}

\newcommand{\MSUPA}{\affiliation{Department of Physics and Astronomy, Michigan State University, East Lansing, MI 48824, USA}}
\newcommand{\MSUCMSE}{\affiliation{Department of Computational Mathematics, Science, and Engineering, Michigan State University, East Lansing, MI 48824, USA}}
\newcommand{\UPitt}{\affiliation{Department of Physics and Astronomy and Pittsburgh Particle Physics, Astrophysics and Cosmology Center (PITT PACC), University of Pittsburgh, 3941 O’Hara Street, Pittsburgh, PA 15260, USA}}

\newcommand{\USzeged}{\affiliation{Department of Experimental Physics, Institute of Physics, University of Szeged, D\'om t\'er 9, 6720 Szeged, Hungary}}
\newcommand{\ELKHSZTE}{\affiliation{HUN-REN--SZTE Stellar Astrophysics Research Group, Szegedi \'ut, Kt. 766, 6500 Baja, Hungary}}
\newcommand{\Konkoly}{\affiliation{Konkoly Observatory, Research Centre for Astronomy and Earth Sciences (CSFK), MTA Center of Excellence, Konkoly-Thege Mikl\'os \'ut 15-17, 1121 Budapest, Hungary}}

\newcommand{\AMNH}{\affiliation{Department of Astrophysics, American Museum of Natural History, Central Park West and 79th Street, New York, NY 10024-5192, USA}}
\newcommand{\UDublin}{\affiliation{School of Physics, Trinity College Dublin, The University of Dublin, Dublin 2, Ireland}}
\newcommand{\VTech}{\affiliation{Department of Physics, Virginia Tech, Blacksburg, VA 24061, USA}}
\newcommand{\USheffield}{\affiliation{Department of Physics and Astronomy, University of Sheffield, Hicks Building, Hounsfield Road, Sheffield S3 7RH, U.K.}}
\newcommand{\Aarhus}{\affiliation{Department of Physics and Astronomy, Aarhus University, Ny Munkegade 120, DK-8000 Aarhus C, Denmark}}

\newcommand{\ICG}{\affiliation{Institute of Cosmology \& Gravitation, University of Portsmouth, Dennis Sciama Building, Burnaby Road, Portsmouth PO1 3FX, UK}}

\newcommand{\Liverpool}{\affiliation{Astrophysics Research Institute, Liverpool John Moores University, Liverpool,  L3 5RF, UK}}
\newcommand{\MaxPlanck}{\affiliation{Max-Planck Institute for Astrophysics, Garching, Germany}}
\newcommand{\GSI}{\affiliation{GSI Helmholtzzentrum f\"ur Schwerionenforschung, Planckstra\ss{}e 1, 64291 Darmstadt, Germany}}
\newcommand{\KyotoU}{\affiliation{Department of Astronomy, Kyoto University, Kitashirakawa-Oiwake-cho, Sakyo-ku, Kyoto, 606-8502. Japan}}
\newcommand{\OSU}{\affiliation{Center for Cosmology and Astroparticle Physics, The Ohio State University, 191 West Woodruff Ave, Columbus, OH 43215, USA}}
\newcommand{\LasCampanas}{\affiliation{Carnegie Observatories, Las Campanas Observatory, Casilla 601, La Serena, Chile}}
\newcommand{\NotreDame}{\affiliation{Department of Physics and Astronomy, University of Notre Dame, Notre Dame, IN 46556, USA}}

\newcommand{\AixMarseille}{\affiliation{Aix Marseille Univ, CNRS, CNES, LAM, Marseille, France}}
\newcommand{\IAP}{\affiliation{Institut d'Astrophysique de Paris, CNRS-Sorbonne Universit\'{e}, 98 bis boulevard Arago, 75014, Paris, France}}
\newcommand{\Thailand}{\affiliation{National Astronomical Research Institute of Thailand, 260 Moo 4, Donkaew, Maerim, Chiang Mai 50180, Thailand}}
\newcommand{\PUC}{\affiliation{Instituto de Astrof\'{i}sica, Facultad de F\'{i}sica, Pontificia Universidad Cat\'{o}lica de Chile, Av. Vicu\~{n}a Mackenna 4860, Santiago, Chile}}
\newcommand{\MIA}{\affiliation{Millennium Institute of Astrophysics, Nuncio Monse\~{n}or S\'{o}tero Sanz 100, Providencia, Santiago, Chile}}
\newcommand{\TAPIR}{\affiliation{TAPIR, Walter Burke Institute for Theoretical Physics, 350-17, Caltech, Pasadena, CA 91125, USA}}
\newcommand{\UTexas}{\affiliation{Department of Astronomy, University of Texas at Austin, Austin, TX, USA}}
\newcommand{\VVS}{\affiliation{Vereniging Voor Sterrenkunde (VVS), Oostmeers 122 C, 8000 Brugge, Belgium}}
\newcommand{\AAVSO}{\affiliation{AAVSO, 185 Alewife Brook Parkway, Suite 410, Cambridge, MA 02138, USA}}
\newcommand{\GEOS}{\affiliation{Groupe Europ\'{e}en d'Observations Stellaires (GEOS), 23 Parc de Levesville, 28300 Bailleau l'Ev\^{e}que, France}}
\newcommand{\BAV}{\affiliation{Bundesdeutsche Arbeitsgemeinschaft für Ver\"{a}nderliche Sterne (BAV), Munsterdamm 90, 12169 Berlin, Germany}}
\newcommand{\LPNHE}{\affiliation{LPNHE, (CNRS/IN2P3, Sorbonne Universit\'{e}, Universit\'{e} Paris Cit\'{e}), Laboratoire de Physique Nucl\'{e}aire et de Hautes \'{E}nergies, 75005, Paris, France}}
\newcommand{\ING}{\affiliation{Isaac Newton Group (ING), Apt. de correos 321, E-38700, Santa Cruz de La Palma, Canary Islands, Spain}}
\newcommand{\ELTE}{\affiliation{ELTE E\"otv\"os Lor\'and University, Institute of Physics and Astronomy, P\'azm\'any P\'eter s\'et\'any 1/A, Budapest, 1117 Hungary}}

%% file: authors.tex
%% first 3 will get shuffled for papers I, II, III
\author[0000-0003-3108-1328]{Lindsey A.\ Kwok}
\Rutgers
\author[0000-0003-2445-3891]{Matthew~R.~Siebert}
\STScI
\author[0000-0001-5975-290X]{Joel Johansson}
\OKC
%% all the rest should be consistent across all three papers
\author[0000-0001-8738-6011]{Saurabh W.\ Jha}
\Rutgers
\author[0000-0002-9388-2932]{St\'{e}phane Blondin}
\AixMarseille
\author[0000-0003-0599-8407]{Luc Dessart}
\IAP
\author[0000-0002-2445-5275]{Ryan J.\ Foley}
\UCSC
\author[0000-0001-5094-8017]{D.\ John Hillier}
\UPitt
\author[0000-0003-2037-4619]{Conor Larison}
\Rutgers
\author[0000-0003-3308-2420]{R\"{u}diger Pakmor}
\MaxPlanck
\author[0000-0001-7380-3144]{Tea Temim}
\Princeton
\author[0000-0003-0123-0062]{Jennifer E.\ Andrews}
\GeminiNorth
\author[0000-0002-4449-9152]{Katie Auchettl}
\Melbourne
\UCSC
\author[0000-0003-3494-343X]{Carles Badenes}
\UPitt
\author[0000-0003-4769-4794]{Barnabas Barna}
\USzeged
\author[0000-0002-4924-444X]{K.\ Azalee Bostroem}
\thanks{LSSTC Catalyst Fellow}
\UA
\author[0000-0002-8092-2077]{Max J.\ B.\ Newman}
\Rutgers
\author[0000-0001-5955-2502]{Thomas G.\ Brink}
\UCB
\author[0000-0003-0416-9818]{Mar\'{i}a Jos\'{e} Bustamante-Rosell}
\UCSC
\author[0000-0002-9830-3880]{Yssavo Camacho-Neves}
\Rutgers
\author[0000-0003-3068-4258]{Alejandro Clocchiatti}
\PUC
\MIA
\author[0000-0003-4263-2228]{David~A.~Coulter}
\UCSC
\author[0000-0002-5680-4660]{Kyle W.\ Davis}
\UCSC
\author[0000-0001-8857-9843]{Maxime Deckers}
\UDublin
\author[0000-0001-9494-179X]{Georgios Dimitriadis}
\UDublin
\author[0000-0002-7937-6371]{Yize Dong}
\UCD
\author[0000-0003-4914-5625]{Joseph Farah}
\LCO
\UCSB
\author[0000-0003-3460-0103]{Alexei V.\ Filippenko}
\UCB
\author[0000-0003-2024-2819]{Andreas Fl\"ors}
\GSI
\author[0000-0003-2238-1572]{Ori D.\ Fox}
\STScI
\author[0000-0003-4069-2817]{Peter Garnavich}
\NotreDame
\author[0000-0003-0209-9246]{Estefania Padilla Gonzalez}
\LCO
\UCSB
%\author[0000-0002-9154-3136]{Melissa L.\ Graham} % opted out
%\DiRAC
\author[0000-0002-4391-6137]{Or Graur}
\ICG
\AMNH
\author[0000-0003-0125-8700]{Franz-Josef Hambsch}
\VVS
\AAVSO
\GEOS
\BAV
\author[0000-0002-0832-2974]{Griffin Hosseinzadeh}
\UA
\author[0000-0003-4253-656X]{D.\ Andrew Howell}
\LCO
\UCSB
\author[0000-0002-8816-6800]{John P.\ Hughes}
\Rutgers
%\author[0000-0002-7612-0469]{Sarah Kendrew}     % opted out
%\ESASTScI
\author[0000-0002-0479-7235]{Wolfgang E.\ Kerzendorf}
\MSUPA
\MSUCMSE
\author[0009-0004-3242-282X]{Xavier K.\ Le Saux}
\UCSC
\author[0000-0003-2611-7269]{Keiichi Maeda}
\KyotoU
\author[0000-0002-9770-3508]{Kate Maguire}
\UDublin
\author[0000-0001-5807-7893]{Curtis McCully}
\LCO
\UCSB
\author[0009-0004-0322-6299]{Cassidy Mihalenko} 
\Melbourne
\ASTROthreeD
\author[0000-0001-9570-0584]{Megan Newsome}
\LCO
\UCSB
\author[0000-0003-3615-9593]{John T.\ O'Brien}
\MSUPA
\author[0000-0002-0744-0047]{Jeniveve Pearson}
\UA
\author[0000-0002-7472-1279]{Craig Pellegrino}
\LCO
\UCSB
\author[0000-0002-2361-7201]{Justin D.\ R.\ Pierel}
\STScI
\author[0000-0002-1633-6495]{Abigail Polin}
\Carnegie
\TAPIR
\author[0000-0002-4410-5387]{Armin Rest}
\STScI
\JHU
\author[0000-0002-7559-315X]{C\'{e}sar Rojas-Bravo}
\UCSC
\author[0000-0003-4102-380X]{David J.\ Sand}
\UA
\author[0009-0002-5096-1689]{Michaela Schwab}
\Rutgers
\author[0000-0002-9301-5302]{Melissa Shahbandeh}
\STScI
\author[0000-0002-4022-1874]{Manisha Shrestha}
\UA
\author[0000-0001-5510-2424]{Nathan Smith}
\UA
\author[0000-0002-7756-4440]{Louis-Gregory Strolger}
\STScI
\author[0000-0003-4610-1117]{Tam\'as Szalai}
\USzeged
\ELKHSZTE
\author[0000-0002-5748-4558]{Kirsty Taggart}
\UCSC
\author[0000-0003-0794-5982]{Giacomo Terreran}
\LCO
\UCSB
\author[0000-0001-9834-3439]{Jacco H.\ Terwel}
\UDublin
\ING
\author[0000-0002-1481-4676]{Samaporn Tinyanont}
\Thailand
\author[0000-0001-8818-0795]{Stefano Valenti}
\UCD
\author[0000-0001-8764-7832]{J\'{o}zsef Vink\'{o}}
\Konkoly
\USzeged
\ELTE
\UTexas
\author[0000-0003-1349-6538]{J.\ Craig Wheeler}
\UTexas
\author[0000-0002-6535-8500]{Yi Yang}
\UCB
\author[0000-0002-2636-6508]{WeiKang Zheng}
\UCB
%
% commented out folks from the Ashall team who haven't yet responded to the
% email; let's make sure everyone has ample opportunity to opt back in!
%
\author[0000-0002-5221-7557]{Chris Ashall}
\VTech
%\author[0000-0001-5393-1608]{E. Baron}
%\UOklahoma
%\Hamburg
%\author[0000-0003-4625-6629]{Chris R.\ Burns}
%\Carnegie
\author[0000-0002-7566-6080]{James M.\ DerKacy}
\VTech
%\author[0000-0001-5888-2542]{Tyco Mera Evans}
%\FSU
%\author[0000-0002-5253-3584]{Alec Fisher}
%\FSU
\author[0000-0002-1296-6887]{Llu\'is Galbany}
\ICE
\IEEC
\author[0000-0002-4338-6586]{Peter Hoeflich}
\FSU
%\author[0000-0003-1039-2928]{Eric Hsiao}
%\FSU
\author[0000-0001-6069-1139]{Thomas de Jaeger}
\LPNHE
%\author[0000-0001-6209-838X]{Emir Karamehmetoglu}
%\Aarhus
%\author[0000-0002-6650-694X]{Kevin Krisciunas}
%\TAMU
%\Mitchell
%\author[0000-0001-8367-7591]{Sahana Kumar}
%\FSU
\author[0000-0002-3900-1452]{Jing Lu}
\MSUPA
\author[0000-0003-0733-7215]{Justyn Maund}
\USheffield
%\author[0000-0001-6876-8284]{Paolo A.\ Mazzali}
%\Liverpool
%\MaxPlanck
\author[0000-0001-7186-105X]{Kyle Medler}
\Liverpool
\author[0000-0003-2535-3091]{Nidia Morrell}
\LasCampanas
%\author[0000-0003-2734-0796]{Mark. M.\ Phillips}
%\LasCampanas
\author[0000-0003-4631-1149]{Benjamin J.\ Shappee} 
\IfA
\author[0000-0002-5571-1833]{Maximilian Stritzinger}
\Aarhus
\author[0000-0002-8102-181X]{Nicholas Suntzeff}
\TAMU
\Mitchell
%\author[0000-0002-0036-9292]{Charles Telesco}
%\UFlorida
\author[0000-0002-2471-8442]{Michael Tucker}
\thanks{CCAPP Fellow}
\OSU
\author[0000-0001-7092-9374]{Lifan Wang}
\TAMU
\Mitchell

%% file: lineid_merger_clumped_7.tex
\begin{table*}[t]
\centering
\footnotesize
\caption{Lines in the wavelength range 0.35-14\,$\mu$m in our modified merger model (MERGER with clumping + 0.1\,\msun\ central mass) at 338 days post explosion whose absolute Sobolev EW exceeds 5\% of the largest absolute EW in the wavelength range 0.35-1\,$\mu$m.}\label{tab:lineid}
\begin{tabular}{rc|rc|rc|rc|rc|rc|rc}
\hline
\multicolumn{1}{c}{$\lambda_\mathrm{air}$} & \multicolumn{1}{c|}{Ion} & \multicolumn{1}{c}{$\lambda_\mathrm{air}$} & \multicolumn{1}{c|}{Ion} & \multicolumn{1}{c}{$\lambda_\mathrm{air}$} & \multicolumn{1}{c|}{Ion} & \multicolumn{1}{c}{$\lambda_\mathrm{air}$} & \multicolumn{1}{c|}{Ion} & \multicolumn{1}{c}{$\lambda_\mathrm{air}$} & \multicolumn{1}{c|}{Ion} & \multicolumn{1}{c}{$\lambda_\mathrm{air}$} & \multicolumn{1}{c|}{Ion} & \multicolumn{1}{c}{$\lambda_\mathrm{air}$} & \multicolumn{1}{c}{Ion} \\
\multicolumn{1}{c}{($\mu$m)} &  & \multicolumn{1}{c}{($\mu$m)} &  & \multicolumn{1}{c}{($\mu$m)} &  & \multicolumn{1}{c}{($\mu$m)} &  & \multicolumn{1}{c}{($\mu$m)} &  & \multicolumn{1}{c}{($\mu$m)} &  & \multicolumn{1}{c}{($\mu$m)} &  \\
\hline
0.427\phantom{\,$\dag$} & Fe\,\sc{i} $^{{\mathrm{{a}}}}$ & 0.501\phantom{\,$\dag$} & [Fe\,\sc{iii}] $^{{\mathrm{{i}}}}$ & 0.717\phantom{\,$\dag$} & [Fe\,\sc{ii}] $^{{\mathrm{{m}}}}$ & 0.923\phantom{\,$\dag$} & [Fe\,\sc{ii}] $^{{\mathrm{{t}}}}$ & 1.321\phantom{\,$\dag$} & [Fe\,\sc{ii}] $^{{\mathrm{{y}}}}$ & 1.954\phantom{\,$\dag$} & [Fe\,\sc{ii}] $^{{\mathrm{{aa}}}}$ & 5.672\phantom{\,$\dag$} & [Fe\,\sc{ii}] $^{{\mathrm{{ah}}}}$ \\
0.429\,$\dag$ & [Fe\,\sc{ii}] $^{{\mathrm{{b}}}}$ & 0.501\phantom{\,$\dag$} & Fe\,\sc{i} $^{{\mathrm{{h}}}}$ & 0.729\,$\dag$ & [Ca\,\sc{ii}] $^{{\mathrm{{n}}}}$ & 0.927\phantom{\,$\dag$} & [Fe\,\sc{ii}] $^{{\mathrm{{t}}}}$ & 1.328\phantom{\,$\dag$} & [Fe\,\sc{ii}] $^{{\mathrm{{y}}}}$ & 1.967\phantom{\,$\dag$} & [Fe\,\sc{ii}] $^{{\mathrm{{ab}}}}$ & 5.703\phantom{\,$\dag$} & [Co\,\sc{ii}] $^{{\mathrm{{ai}}}}$ \\
0.431\phantom{\,$\dag$} & Fe\,\sc{i} $^{{\mathrm{{a}}}}$ & 0.502\phantom{\,$\dag$} & Fe\,\sc{ii} $^{{\mathrm{{g}}}}$ & 0.732\,$\dag$ & [Ca\,\sc{ii}] $^{{\mathrm{{n}}}}$ & 0.934\,$\dag$ & [Co\,\sc{ii}] $^{{\mathrm{{v}}}}$ & 1.372\phantom{\,$\dag$} & [Fe\,\sc{ii}] $^{{\mathrm{{y}}}}$ & 2.007\phantom{\,$\dag$} & [Fe\,\sc{ii}] $^{{\mathrm{{ad}}}}$ & 6.212\phantom{\,$\dag$} & [Co\,\sc{ii}] $^{{\mathrm{{ai}}}}$ \\
0.433\phantom{\,$\dag$} & Fe\,\sc{i} $^{{\mathrm{{a}}}}$ & 0.505\phantom{\,$\dag$} & Fe\,\sc{i} $^{{\mathrm{{h}}}}$ & 0.738\,$\dag$ & [Ni\,\sc{ii}] $^{{\mathrm{{o}}}}$ & 0.934\phantom{\,$\dag$} & [Co\,\sc{ii}] $^{{\mathrm{{w}}}}$ & 1.497\phantom{\,$\dag$} & [Co\,\sc{ii}] $^{{\mathrm{{z}}}}$ & 2.015\phantom{\,$\dag$} & [Fe\,\sc{ii}] $^{{\mathrm{{ae}}}}$ & 6.634\,$\dag$ & [Ni\,\sc{ii}]\phantom{ $^{\mathrm{a}}$} \\
0.436\phantom{\,$\dag$} & [Fe\,\sc{ii}] $^{{\mathrm{{b}}}}$ & 0.508\phantom{\,$\dag$} & Fe\,\sc{i} $^{{\mathrm{{h}}}}$ & 0.739\phantom{\,$\dag$} & [Fe\,\sc{ii}] $^{{\mathrm{{m}}}}$ & 0.953\phantom{\,$\dag$} & [S\,\sc{iii}] $^{{\mathrm{{u}}}}$ & 1.533\phantom{\,$\dag$} & [Fe\,\sc{ii}] $^{{\mathrm{{aa}}}}$ & 2.046\phantom{\,$\dag$} & [Fe\,\sc{ii}] $^{{\mathrm{{ad}}}}$ & 6.719\phantom{\,$\dag$} & [Fe\,\sc{ii}] $^{{\mathrm{{ah}}}}$ \\
0.438\,$\dag$ & Fe\,\sc{i}] $^{{\mathrm{{c}}}}$ & 0.516\phantom{\,$\dag$} & [Fe\,\sc{ii}]\phantom{ $^{\mathrm{a}}$} & 0.741\phantom{\,$\dag$} & [Ni\,\sc{ii}] $^{{\mathrm{{o}}}}$ & 0.964\phantom{\,$\dag$} & [Co\,\sc{ii}] $^{{\mathrm{{v}}}}$ & 1.547\phantom{\,$\dag$} & [Co\,\sc{ii}] $^{{\mathrm{{z}}}}$ & 2.133\phantom{\,$\dag$} & [Fe\,\sc{ii}] $^{{\mathrm{{ad}}}}$ & 6.918\phantom{\,$\dag$} & [Ni\,\sc{ii}]\phantom{ $^{\mathrm{a}}$} \\
0.438\phantom{\,$\dag$} & Fe\,\sc{i}] $^{{\mathrm{{d}}}}$ & 0.517\phantom{\,$\dag$} & Fe\,\sc{i} $^{{\mathrm{{j}}}}$ & 0.745\phantom{\,$\dag$} & [Fe\,\sc{ii}] $^{{\mathrm{{m}}}}$ & 0.994\phantom{\,$\dag$} & [Co\,\sc{ii}] $^{{\mathrm{{w}}}}$ & 1.599\phantom{\,$\dag$} & [Fe\,\sc{ii}] $^{{\mathrm{{aa}}}}$ & 2.206\phantom{\,$\dag$} & Na\,\sc{i} $^{{\mathrm{{af}}}}$ & 6.983\,$\dag$ & [Ar\,\sc{ii}]\phantom{ $^{\mathrm{a}}$} \\
0.440\phantom{\,$\dag$} & Fe\,\sc{i}] $^{{\mathrm{{d}}}}$ & 0.517\phantom{\,$\dag$} & Fe\,\sc{ii} $^{{\mathrm{{g}}}}$ & 0.764\phantom{\,$\dag$} & [Fe\,\sc{ii}] $^{{\mathrm{{p}}}}$ & 1.019\,$\dag$ & [Co\,\sc{ii}] $^{{\mathrm{{v}}}}$ & 1.626\phantom{\,$\dag$} & [Co\,\sc{ii}] $^{{\mathrm{{z}}}}$ & 2.208\phantom{\,$\dag$} & Na\,\sc{i} $^{{\mathrm{{af}}}}$ & 7.347\,$\dag$ & [Ni\,\sc{iii}] $^{{\mathrm{{aj}}}}$ \\
0.441\phantom{\,$\dag$} & [Fe\,\sc{ii}] $^{{\mathrm{{b}}}}$ & 0.517\phantom{\,$\dag$} & Fe\,\sc{i}\phantom{ $^{\mathrm{a}}$} & 0.769\phantom{\,$\dag$} & [Fe\,\sc{ii}] $^{{\mathrm{{p}}}}$ & 1.025\phantom{\,$\dag$} & [Co\,\sc{ii}] $^{{\mathrm{{v}}}}$ & 1.634\phantom{\,$\dag$} & [Co\,\sc{ii}] $^{{\mathrm{{z}}}}$ & 2.224\phantom{\,$\dag$} & [Fe\,\sc{ii}] $^{{\mathrm{{ae}}}}$ & 7.505\,$\dag$ & [Ni\,\sc{i}] $^{{\mathrm{{ak}}}}$ \\
0.442\phantom{\,$\dag$} & Fe\,\sc{i}] $^{{\mathrm{{d}}}}$ & 0.523\phantom{\,$\dag$} & Fe\,\sc{i} $^{{\mathrm{{j}}}}$ & 0.789\phantom{\,$\dag$} & [Ni\,\sc{iii}] $^{{\mathrm{{q}}}}$ & 1.028\phantom{\,$\dag$} & [Co\,\sc{ii}] $^{{\mathrm{{v}}}}$ & 1.644\phantom{\,$\dag$} & [Fe\,\sc{ii}] $^{{\mathrm{{aa}}}}$ & 2.244\phantom{\,$\dag$} & [Fe\,\sc{ii}] $^{{\mathrm{{ad}}}}$ & 7.788\phantom{\,$\dag$} & [Fe\,\sc{iii}]\phantom{ $^{\mathrm{a}}$} \\
0.442\,$\dag$ & [Fe\,\sc{ii}]\phantom{ $^{\mathrm{a}}$} & 0.527\phantom{\,$\dag$} & Fe\,\sc{i} $^{{\mathrm{{k}}}}$ & 0.833\phantom{\,$\dag$} & Fe\,\sc{i} $^{{\mathrm{{r}}}}$ & 1.032\phantom{\,$\dag$} & [S\,\sc{ii}]\phantom{ $^{\mathrm{a}}$} & 1.664\phantom{\,$\dag$} & [Fe\,\sc{ii}] $^{{\mathrm{{aa}}}}$ & 2.308\phantom{\,$\dag$} & [Ni\,\sc{ii}] $^{{\mathrm{{ac}}}}$ & 8.403\,$\dag$ & [Ni\,\sc{iv}] $^{{\mathrm{{al}}}}$ \\
0.443\phantom{\,$\dag$} & Fe\,\sc{i}] $^{{\mathrm{{c}}}}$ & 0.527\phantom{\,$\dag$} & [Fe\,\sc{iii}] $^{{\mathrm{{i}}}}$ & 0.839\phantom{\,$\dag$} & Fe\,\sc{i} $^{{\mathrm{{r}}}}$ & 1.097\phantom{\,$\dag$} & [Co\,\sc{ii}] $^{{\mathrm{{v}}}}$ & 1.677\phantom{\,$\dag$} & [Fe\,\sc{ii}] $^{{\mathrm{{aa}}}}$ & 2.369\phantom{\,$\dag$} & [Ni\,\sc{ii}] $^{{\mathrm{{ac}}}}$ & 8.989\,$\dag$ & [Ar\,\sc{iii}]\phantom{ $^{\mathrm{a}}$} \\
0.445\phantom{\,$\dag$} & [Fe\,\sc{ii}] $^{{\mathrm{{b}}}}$ & 0.533\phantom{\,$\dag$} & Fe\,\sc{i} $^{{\mathrm{{k}}}}$ & 0.850\phantom{\,$\dag$} & Ca\,\sc{ii} $^{{\mathrm{{s}}}}$ & 1.128\phantom{\,$\dag$} & [Co\,\sc{ii}] $^{{\mathrm{{v}}}}$ & 1.711\phantom{\,$\dag$} & [Fe\,\sc{ii}] $^{{\mathrm{{aa}}}}$ & 2.911\phantom{\,$\dag$} & [Ni\,\sc{ii}] $^{{\mathrm{{ac}}}}$ & 10.508\,$\dag$ & [S\,\sc{iv}]\phantom{ $^{\mathrm{a}}$} \\
0.446\phantom{\,$\dag$} & Fe\,\sc{i}] $^{{\mathrm{{c}}}}$ & 0.537\phantom{\,$\dag$} & Fe\,\sc{i} $^{{\mathrm{{k}}}}$ & 0.850\phantom{\,$\dag$} & [Ni\,\sc{iii}] $^{{\mathrm{{q}}}}$ & 1.161\phantom{\,$\dag$} & Fe\,\sc{i} $^{{\mathrm{{x}}}}$ & 1.728\phantom{\,$\dag$} & [Co\,\sc{ii}] $^{{\mathrm{{z}}}}$ & 3.119\phantom{\,$\dag$} & [Ni\,\sc{i}]\phantom{ $^{\mathrm{a}}$} & 10.520\,$\dag$ & [Co\,\sc{ii}] $^{{\mathrm{{v}}}}$ \\
0.448\phantom{\,$\dag$} & Fe\,\sc{i}] $^{{\mathrm{{c}}}}$ & 0.540\phantom{\,$\dag$} & Fe\,\sc{i} $^{{\mathrm{{k}}}}$ & 0.854\phantom{\,$\dag$} & Ca\,\sc{ii} $^{{\mathrm{{s}}}}$ & 1.169\phantom{\,$\dag$} & Fe\,\sc{i} $^{{\mathrm{{x}}}}$ & 1.736\phantom{\,$\dag$} & [Co\,\sc{ii}] $^{{\mathrm{{z}}}}$ & 3.206\,$\dag$ & [Ca\,\sc{iv}]\phantom{ $^{\mathrm{a}}$} & 10.679\phantom{\,$\dag$} & [Ni\,\sc{ii}] $^{{\mathrm{{am}}}}$ \\
0.455\phantom{\,$\dag$} & Fe\,\sc{ii} $^{{\mathrm{{e}}}}$ & 0.541\phantom{\,$\dag$} & Fe\,\sc{i} $^{{\mathrm{{k}}}}$ & 0.862\phantom{\,$\dag$} & [Fe\,\sc{ii}] $^{{\mathrm{{t}}}}$ & 1.188\phantom{\,$\dag$} & Fe\,\sc{i} $^{{\mathrm{{x}}}}$ & 1.745\phantom{\,$\dag$} & [Fe\,\sc{ii}] $^{{\mathrm{{aa}}}}$ & 3.393\phantom{\,$\dag$} & [Ni\,\sc{iii}] $^{{\mathrm{{ag}}}}$ & 10.999\phantom{\,$\dag$} & [Ni\,\sc{iii}] $^{{\mathrm{{aj}}}}$ \\
0.458\phantom{\,$\dag$} & Fe\,\sc{ii} $^{{\mathrm{{e}}}}$ & 0.543\phantom{\,$\dag$} & Fe\,\sc{i} $^{{\mathrm{{k}}}}$ & 0.866\phantom{\,$\dag$} & Fe\,\sc{i} $^{{\mathrm{{r}}}}$ & 1.188\phantom{\,$\dag$} & Fe\,\sc{i} $^{{\mathrm{{x}}}}$ & 1.748\phantom{\,$\dag$} & [Fe\,\sc{ii}] $^{{\mathrm{{ab}}}}$ & 3.801\phantom{\,$\dag$} & [Ni\,\sc{iii}] $^{{\mathrm{{ag}}}}$ & 11.164\phantom{\,$\dag$} & [Co\,\sc{ii}] $^{{\mathrm{{v}}}}$ \\
0.466\,$\dag$ & [Fe\,\sc{iii}] $^{{\mathrm{{f}}}}$ & 0.543\phantom{\,$\dag$} & Fe\,\sc{i} $^{{\mathrm{{k}}}}$ & 0.866\phantom{\,$\dag$} & Ca\,\sc{ii} $^{{\mathrm{{s}}}}$ & 1.197\phantom{\,$\dag$} & Fe\,\sc{i} $^{{\mathrm{{x}}}}$ & 1.797\phantom{\,$\dag$} & [Fe\,\sc{ii}] $^{{\mathrm{{aa}}}}$ & 4.075\phantom{\,$\dag$} & [Fe\,\sc{ii}] $^{{\mathrm{{ah}}}}$ & 11.304\phantom{\,$\dag$} & [Ni\,\sc{i}] $^{{\mathrm{{ak}}}}$ \\
0.470\phantom{\,$\dag$} & [Fe\,\sc{iii}] $^{{\mathrm{{f}}}}$ & 0.545\phantom{\,$\dag$} & Fe\,\sc{i} $^{{\mathrm{{k}}}}$ & 0.869\phantom{\,$\dag$} & Fe\,\sc{i} $^{{\mathrm{{r}}}}$ & 1.249\phantom{\,$\dag$} & [Fe\,\sc{ii}] $^{{\mathrm{{y}}}}$ & 1.800\phantom{\,$\dag$} & [Fe\,\sc{ii}] $^{{\mathrm{{aa}}}}$ & 4.114\,$\dag$ & [Fe\,\sc{ii}] $^{{\mathrm{{ah}}}}$ & 11.723\phantom{\,$\dag$} & [Ni\,\sc{iv}] $^{{\mathrm{{al}}}}$ \\
0.481\phantom{\,$\dag$} & [Fe\,\sc{ii}]\phantom{ $^{\mathrm{a}}$} & 0.546\phantom{\,$\dag$} & Fe\,\sc{i} $^{{\mathrm{{k}}}}$ & 0.882\phantom{\,$\dag$} & Fe\,\sc{i} $^{{\mathrm{{r}}}}$ & 1.257\,$\dag$ & [Fe\,\sc{ii}] $^{{\mathrm{{y}}}}$ & 1.809\phantom{\,$\dag$} & [Fe\,\sc{ii}] $^{{\mathrm{{aa}}}}$ & 4.434\phantom{\,$\dag$} & [Fe\,\sc{ii}] $^{{\mathrm{{ah}}}}$ & 11.885\,$\dag$ & [Co\,\sc{iii}]\phantom{ $^{\mathrm{a}}$} \\
0.489\phantom{\,$\dag$} & [Fe\,\sc{ii}]\phantom{ $^{\mathrm{a}}$} & 0.589\,$\dag$ & Na\,\sc{i} $^{{\mathrm{{l}}}}$ & 0.889\phantom{\,$\dag$} & [Fe\,\sc{ii}] $^{{\mathrm{{t}}}}$ & 1.270\phantom{\,$\dag$} & [Fe\,\sc{ii}] $^{{\mathrm{{y}}}}$ & 1.811\phantom{\,$\dag$} & [Fe\,\sc{ii}] $^{{\mathrm{{ab}}}}$ & 4.606\phantom{\,$\dag$} & [Fe\,\sc{ii}] $^{{\mathrm{{ah}}}}$ & 11.998\phantom{\,$\dag$} & [Ni\,\sc{i}]\phantom{ $^{\mathrm{a}}$} \\
0.492\phantom{\,$\dag$} & Fe\,\sc{ii} $^{{\mathrm{{g}}}}$ & 0.590\,$\dag$ & Na\,\sc{i} $^{{\mathrm{{l}}}}$ & 0.903\phantom{\,$\dag$} & [Fe\,\sc{ii}] $^{{\mathrm{{t}}}}$ & 1.279\phantom{\,$\dag$} & [Fe\,\sc{ii}] $^{{\mathrm{{y}}}}$ & 1.895\phantom{\,$\dag$} & [Fe\,\sc{ii}] $^{{\mathrm{{aa}}}}$ & 4.888\phantom{\,$\dag$} & [Fe\,\sc{ii}] $^{{\mathrm{{ah}}}}$ & 12.252\,$\dag$ & [Co\,\sc{i}]\phantom{ $^{\mathrm{a}}$} \\
0.494\phantom{\,$\dag$} & Fe\,\sc{i} $^{{\mathrm{{h}}}}$ & 0.630\,$\dag$ & [O\,\sc{i}]\phantom{ $^{\mathrm{a}}$} & 0.905\phantom{\,$\dag$} & [Fe\,\sc{ii}] $^{{\mathrm{{t}}}}$ & 1.294\phantom{\,$\dag$} & [Fe\,\sc{ii}] $^{{\mathrm{{y}}}}$ & 1.903\phantom{\,$\dag$} & [Co\,\sc{ii}] $^{{\mathrm{{z}}}}$ & 5.061\phantom{\,$\dag$} & [Fe\,\sc{ii}] $^{{\mathrm{{ah}}}}$ & 12.725\phantom{\,$\dag$} & [Ni\,\sc{ii}] $^{{\mathrm{{am}}}}$ \\
0.499\phantom{\,$\dag$} & Fe\,\sc{i} $^{{\mathrm{{h}}}}$ & 0.716\phantom{\,$\dag$} & [Fe\,\sc{ii}] $^{{\mathrm{{m}}}}$ & 0.907\phantom{\,$\dag$} & [S\,\sc{iii}] $^{{\mathrm{{u}}}}$ & 1.298\phantom{\,$\dag$} & [Fe\,\sc{ii}] $^{{\mathrm{{y}}}}$ & 1.939\phantom{\,$\dag$} & [Ni\,\sc{ii}] $^{{\mathrm{{ac}}}}$ & 5.339\,$\dag$ & [Fe\,\sc{ii}] $^{{\mathrm{{ah}}}}$ & 
12.811\,$\dag$ & [Ne\,\sc{ii}]\phantom{ $^{\mathrm{a}}$} \\

\hline
\end{tabular}
\flushleft
{\bf Notes:} All wavelengths are given in air. Forbidden and semiforbidden transitions are noted using the appropriate brackets around the ion name. Wavelengths marked with a `$\dag$' symbol denote transitions connected to the ground state. Ions with the same superscript correspond to transitions within the same multiplet.
\end{table*}